%% file: WL-comb.tex
\documentclass[11pt]{article}
\usepackage[utf8]{inputenc}
\usepackage[left=1in,right=1in]{geometry}
\usepackage{authblk}
\usepackage[inline]{showlabels}
\usepackage[pdfstartview=FitH,pdfpagemode=None]{hyperref}
\usepackage{amsmath}
\usepackage{amsfonts}
\usepackage{cite}
\usepackage{tabularx}

\title{Exact 1/N expansion of Wilson loop correlators in $\mathcal{N}=4$ Super-Yang-Mills theory}
\author[1,2]{Wolfgang M\"uck \thanks{E-mail: \texttt{mueck@na.infn.it}}}
\affil[1]{Dipartimento di Fisica ``Ettore Pancini", Universit\`a degli Studi di Napoli ``Federico II" \authorcr Via Cintia, 80126 Napoli, Italy}
\affil[2]{Istituto Nazionale di Fisica Nucleare, Sezione di Napoli \authorcr Via Cintia, 80126 Napoli, Italy}
\date{}
\input{Definitions}

\begin{document}
\maketitle
\begin{abstract}
Supersymmetric circular Wilson loops in $\mathcal{N}=4$ Super-Yang-Mills theory are discussed starting from their Gaussian matrix model representations. Previous results on the generating functions of Wilson loops are reviewed and extended to the more general case of two different loop contours, which is necessary to discuss coincident loops with opposite orientations. A combinatorial formula representing the connected correlators of multiply wound Wilson loops in terms of the matrix model solution is derived. Two new results are obtained on the expectation value of the circular Wilson loop, the expansion of which into a series in $1/N$ and to all orders in the 't~Hooft coupling $\lambda$ was derived by Drukker and Gross about twenty years ago. The connected correlators of two multiply wound Wilson loops with arbitrary winding numbers are calculated as a series in $1/N$. The coefficient functions are derived not only as power series in $\lambda$, but also to all orders in $\lambda$ by expressing them in terms of the coefficients of the Drukker and Gross series. This provides an efficient way to calculate the $1/N$ series, which can probably be generalized to higher-point correlators.
\end{abstract}
%%%%%%%%%%%%%%%%%%%%%%%%%%%%%%%%%%%%%%%%%%%%
\input{intro}
%%%%%%%%%%%%%%%%%%%%%%%%%%%%%%%%%%%%%%%%%%%%
\input{WLcorr}

\input{N4corr}

\input{exact}
\input{fundamental}

\input{two-point} 
%%%%%%%%%%%%%%%%%%%%%%%%%%%%%%%%%%%%%%%%%%%%
\input{conc}

%%%%%%%%%%%%%%%%%%%%%%%%%%%%%%%%%%%%%%%%%%%%
%
\section*{Acknowledgements}
This work was supported in part by  the INFN, research initiative STEFI.
%%%%%%%%%%%%%%%%%%%%%%%%%%%%%%%%%%%%%%%%%%%%
\begin{appendix}
\input{appS}

\input{apphyp}
\end{appendix}
%%%%%%%%%%%%%%%%%%%%%%%%%%%%%%%%%%%%%%%%%%%%
\bibliographystyle{JHEPnotes}
\bibliography{MM}
\end{document}

%% file: Definitions.tex
\numberwithin{equation}{section}

\newcommand{\ie}{i.e.,\ }

\newcommand{\rmd}{\,\mathrm{d}}

\newcommand{\Tr}{\operatorname{Tr}}

\newcommand{\re}{\operatorname{Re}}

\newcommand{\e}[1]{\operatorname{e}^{#1}}

\newcommand{\genhypF}[3]{{}_{#1}\mathrm{F}_{#2}\left(#3\right)}
\newcommand{\hypF}[1]{\genhypF{2}{1}{#1}}

\newcommand{\Laguerre}{\operatorname{L}}

\newcommand{\WhitM}{\operatorname{M}}

\newcommand{\BesselI}[1][0]{\operatorname{I}_{#1}}

\newcommand{\vev}[1]{\left\langle #1 \right\rangle}
\newcommand{\vvev}[1]{\vev{\mspace{-3mu}\vev{#1}\mspace{-3mu}}}

\newcommand{\Order}{\operatorname{\mathcal{O}}}

\newcommand{\bra}[1]{\langle #1 |}
\newcommand{\ket}[1]{| #1 \rangle}

\newcommand{\Dk}[1]{\mathcal{D}^{(k_1,k_2)}_{#1}}
\newcommand{\tDk}[1]{\tilde{\mathcal{D}}^{(k_1,k_2)}_{#1}}

\newcommand{\Wprod}{\mathcal{W}^{(k_1)}\mathcal{W}^{(k_2)}}

%% file: intro.tex
\section{Introduction}

Wilson loops play an important role in testing the AdS/CFT correspondence \cite{Maldacena:1998im, Rey:1998ik, Drukker:1999zq} beyond the planar approximation \cite{tHooft:1973alw, Brezin:1977sv, Itzykson:1979fi}. Amongst the simplest cases one finds the class of $\frac12$-BPS circular Wilson loops and their correlators in $\mathcal{N}=4$ Super-Yang-Mills (SYM) theory with gauge group U$(N)$ or SU$(N)$.
On the one hand, the holographic dual configurations of strings or D-branes fully capture the planar approximation in the limit of large 't~Hooft coupling $\lambda$ \cite{Drukker:2005kx, Yamaguchi:2006te, Yamaguchi:2006tq, Gomis:2006sb, Lunin:2006xr, Gomis:2006im}, and a lot of effort has been dedicated to obtain corrections in $1/\lambda$ \cite{Forste:1999qn, Drukker:2000ep, Semenoff:2001xp, Kruczenski:2008zk, Faraggi:2011bb, Faraggi:2011ge, Kristjansen:2012nz, Faraggi:2014tna, Faraggi:2016ekd, Horikoshi:2016hds, Forini:2017whz, Aguilera-Damia:2018twq}. 
On the other hand, localization \cite{Pestun:2007rz, Pestun:2016zxk, Zarembo:2016bbk} maps the calculation of these Wilson loop correlators to the solution of a Gaussian matrix model \cite{Erickson:2000af, Drukker:2000rr, Akemann:2001st, Hartnoll:2006is, Fiol:2013hna}, which, in principle, is exact in both, $\lambda$ and $N$, although extracting the $1/N$ expansion in a useful form is not easy. A standard method to construct the $1/N$ expansion is topological recursion or the loop equation approach \cite{Ambjorn:1992gw, Okuyama:2006jc, Chen-Lin:2016kkk, Gordon:2017dvy, Okuyama:2017feo, Okuyama:2018aij, Beccaria:2021alk}. Colour invariants have also been used \cite{Fiol:2018yuc} or more direct approaches \cite{Drukker:2000rr, CanazasGaray:2018cpk}. More recently, the $1/N$ expansion of correlators involving Wilson loops has been studied with the help of the Toda integrability structure \cite{Beccaria:2020ykg}, with a particular emphasis on the strong coupling behaviour \cite{Giombi:2018hsx}. 

A useful approach to Wilson loops is to consider suitable generating functions. Quite generally, Wilson loop generating functions are most elegantly formulated in the language of symmetric functions \cite{Marino:2005sj}, which allows to encode the information on Wilson loops in arbitrary representations of the gauge group and to translate between different sets of basis correlators using combinatorial identities. The generating functions for higher rank Wilson loops introduced in \cite{Hartnoll:2006is} are contained in this language as special cases. In particular, the connected correlators of multiply would Wilson loops turn out to be a natural basis to work with and are a key ingredient in the proof of an interesting involution property \cite{Okuyama:2018aij, Fiol:2018yuc, CanazasGaray:2019mgq, Muck:2019hnz}. In the case of $\mathcal{N}=4$ SYM theory with gauge group U$(N)$, the expression of these correlators in terms of the matrix model solution has been worked out in \cite{Okuyama:2018aij, CanazasGaray:2019mgq, Muck:2019hnz}, and it will be one of the aims of this paper to further elaborate on this relation. 

At a time when localization had not been established yet as a theorem \cite{Pestun:2007rz} and the relation between supersymmetric Wilson loops in $\mathcal{N}=4$ SYM theory and the Gaussian matrix model had only been conjectured \cite{Erickson:2000af}, Drukker and Gross  \cite{Drukker:2000rr} calculated the circular Wilson loop expectation value exactly, as a series in $1/N$ and to all orders in $\lambda$. To date, their result remains a rare example of a full series in $1/N$ that can be obtained from the exact matrix model solution. 

The aim of this paper is to develop methods that allow to extract as many results as possible about coincident circular Wilson loops from the exact matrix model expression of the general Wilson loop generating functions. In particular, we shall be interested in the correlators of multiply wound Wilson loops,
\begin{equation}
\label{intro:WLcorr}
	\mathcal{W}^{(k_1,k_2,\ldots,k_h)} = \vev{\prod_{i=1}^h \Tr U^{k_i}}
\end{equation}
and their connected variants, where the non-zero integers $k_1, k_2, \ldots k_h$ represent the winding numbers. Following Okuyama's slight abuse of nomenclature \cite{Okuyama:2018aij}, we will also call such correlators ``$h$-point functions''. 

The structure of the paper is as follows. We will start by reviewing, in section~\ref{WLcorr}, the general theory of Wilson loop generating functions in the language of symmetric functions \cite{Marino:2005sj, Muck:2019hnz}. This review will end with a generalization to the case of two independent contours, which is necessary for treating coincident circular loops winding with opposite orientations. 
In section~\ref{N4}, we discuss these Wilson loop generating functions in the case of $\mathcal{N}=4$ SYM theory, in which they can be obtained exactly by solving a Gaussian matrix model. The result of this discussion will be a general formula for the connected correlators of multiply wound Wilson loops in terms of the traces of symmetrized matrix products, generalizing the results of \cite{Okuyama:2018aij, Muck:2019hnz}.
Section~\ref{exact} is dedicated to the manipulation of the matrix model result using harmonic oscillator algebra. The deep relation between the Gaussian matrix model and the harmonic oscillator is well known, but the very elegant treatment by Okuyama \cite{Okuyama:2018yep} has gone nearly unnoticed. Therefore, we shall review it here.
In section~\ref{onept}, we will consider the one-point functions $\mathcal{W}^{(k_1)}$ starting with a review of Drukker and Gross' exact series in $1/N$. We will be able to add two new results here. First, we find a recursive set of differential equations, from which the series can be constructed and, second, we provide a new derivation of the series in $1/N$, which results in an explicit combinatorial formula for the numerical coefficients in this series that have been defined only recursively. 
These two results find their analogues in the treatment of the connected two-point functions, which is carried out in section~\ref{tp}. The direct approach will result in the full series in $1/N$, but the individual terms are given only as series in $\lambda$. It will be checked that the leading term in $1/N$ coincides with known results. Using the differential equation approach, however, we will show that the series can be constructed from the knowledge of the Drukker and Gross' series for the one-point functions, resulting in expressions to all order in $\lambda$. A procedure for this construction, which can be coded on the computer, will be given. 
Finally, we will conclude in section~\ref{conc} and add two appendices for technical details. 

%% file: WLcorr.tex
\section{Generating functions of Wilson loops and correlators}
\label{WLcorr}

\subsection{Brief review of combinatorics and symmetric functions}
\label{WLcorr:review}

In this subsection, we will recall some basic combinatorial notions and introduce the symmetric functions. Readers not familiar with them should consult a standard reference such as \cite{Macdonald:1995} or the lecture notes \cite{Lascoux}. 

A \emph{partition} $\lambda \vdash n$ is a weakly decreasing (or weakly increasing) set of \emph{positive} integers $\lambda_i$ ($i=1,2,\ldots$) such that $\sum_i \lambda_i \equiv |\lambda|= n$. The numbers $\lambda_i$ are called the parts of $\lambda$, and the number of the parts of $\lambda$ is denoted by $l(\lambda)$. Obviously, it holds that $l(\lambda)\leq |\lambda|$. 
In order to avoid ambiguities, we will assume $l(\lambda)>0$ throughout the paper, \ie we exclude the empty partition. 
$\lambda$ can be represented graphically as a Young diagram containing columns of lengths $\lambda_1, \lambda_2, \ldots$. Flipping the diagram along its diagonal defines the transpose partition $\lambda^\dagger$. 

Sometimes, the notation 
\begin{equation}
\label{comb:partition}
	\lambda =\prod_i i^{a_i}
\end{equation}
is used, meaning that the integer $i$ is contained in $\lambda$ $a_i$ times. Then, we have
\begin{equation}
\label{comb:part.nums}
	l(\lambda) = \sum\limits_i a_i~,\qquad |\lambda| = \sum\limits_i i\, a_i~.
\end{equation}
The notation \eqref{comb:partition} is particularly useful when relating partitions to permutations. A partition $\lambda$ is associated with the cycle type of a permutation, if the permutation contains $a_i$ cycles of length $i$. Thus, $\lambda$ defines a conjugacy class $C_\lambda$ of the permutation group $\mathfrak{S}_n$. Defining the centralizer size by 
\begin{equation}
\label{Intro:z}
	z_\lambda = \prod\limits_i (a_i!\, i^{a_i})~,
\end{equation}
we have that $|C_\lambda| = |\lambda|!/z_\lambda$ is the size of the conjugacy class, \ie the number of permutations of cycle type $\lambda$.

A \emph{composition} $K$ is a sequence of \emph{positive} integers, $K=(k_1,k_2, \ldots)$, which are called the parts of $K$. The length $l(K)$ is the number of the parts of $K$, and the weight of $K$ is the sum of its parts, $|K|=\sum_i k_i$. Writing the parts of $K$ in weakly decreasing (or increasing) order uniquely associates $K$ with a partition. Therefore, two sequences of positive integers that differ in the order of their parts define different compositions, while they are considered to define the same partition. As for the partitions, we will assume $l(K)>0$.

Given a set $X$, a \emph{set partition} $\nu = (\nu_1, \nu_2,\ldots)$ of $X$ is a sequence of \emph{disjoint} sets $\nu_i$, called the parts of $\nu$, the union of which is $X$. The length $l(\nu)$ is the number of its parts. Of particular interest in this paper are the set partitions of $X=[n]$, where $n$ is some positive integer, and $[n]=(1,2,\ldots,n)$. In particular, given a composition $K$ of weight $|K|$ and length $l(K)$, let $\nu$ be a set partition of $[l(K)]$. This implies that $l(\nu)\leq l(K)$. Then, let $K_\nu$ be defined as the following composition,
\begin{equation}
\label{Intro:K.nu.def}
	K_\nu = \left( \sum\limits_{i\in \nu_1} k_i, \sum\limits_{i\in \nu_2} k_i, \ldots,\sum\limits_{i\in \nu_{l(\nu)}} k_i \right)~.
\end{equation}
Clearly, $|K_\nu|=|K|$ and $l(K_\nu) = l(\nu)$. 

Let us now introduce the symmetric functions \cite{Macdonald:1995,Lascoux}. Let $e_n$, $h_n$, and $p_n$ be the elementary, complete homogeneous and power-sum polynomials of degree $n$, respectively. For $a \in \{e,h,p\}$, given a composition $K$, we define $a_K = \prod_i a_{k_i}$. Because a composition $K$ is uniquely associated with a partition $\lambda$, we can identify $a_\lambda=a_K$.\footnote{Most often, these functions are defined with reference to a partition. For commuting variables, the two definitions are clearly equivalent. For non-commuting variables \cite{Gelfand:1994hg}, using compositions is more appropriate.} These functions form bases of symmetric functions (on some countably infinite alphabet). There are three additional classical bases, the monomials, $m_\lambda$, the Schur basis, $s_\lambda$, and the ``forgotten'' basis, $f_\lambda$. Their role is captured best by considering the Hall inner product, $\vev{\cdot,\cdot}$, or the Cauchy kernel. The monomial basis is the adjoint of the complete homogeneous basis, $\vev{m_\lambda,h_\nu}=\delta_{\lambda\nu}$, the forgotten basis is the adjoint of the elementary basis, $\vev{f_\lambda,e_\nu}=\delta_{\lambda\nu}$, whereas the power-sum basis and the Schur basis satisfy $\vev{p_\lambda,p_\nu}=z_\lambda\delta_{\lambda\nu}$ and $\vev{s_\lambda,s_\nu}=\delta_{\lambda\nu}$, respectively. The Schur functions are related to the monomials by the Kostka matrix \cite{Macdonald:1995}.\footnote{The Kostka matrix was used in \cite{Fiol:2013hna} to obtain the Wilson loops in irreducible representations (Schur basis) from the matrix model solution (monomial basis), but we will not use it here.}

\subsection{Wilson loop generating functions}
\label{WLcorr:genfunc}

Generating functions for Wilson loops in arbitrary representations of the gauge group can be formulated elegantly \cite{Marino:2005sj, Muck:2019hnz} using the language of symmetric functions. We will restrict our treatment to unitary gauge groups. 

Let $U$ be the holonomy of the gauge connection for a single Wilson loop, an ``open'' Wilson loop, so to say. We can take $U$ diagonal, $U=\operatorname{diag}(u_1,u_2,\ldots)$ and denote by $u= (u_1,u_2,\ldots)$ the alphabet of its eigenvalues.\footnote{We will formally consider a countably infinite set of diagonal entries, almost all of which are zero.} 
Then, it is obvious that the $n$-fold multiply-wound Wilson loop
\begin{equation}
\label{WL:Un.pn}
	\Tr U^n = \sum_i u_i^n = p_n(u)
\end{equation}
is simply given by the power-sum symmetric polynomial of degree $n$ in the eigenvalues. Introduce an alphabet of real parameters $y = (y_1,y_2,\ldots)$ and define the two generating functions\footnote{$H(y)$ is the so-called Cauchy kernel. It is also known as the Ooguri-Vafa operator \cite{Ooguri:1999bv}.}
\begin{equation}
\label{WL:EH}
	E(y) = \prod\limits_{i,j} (1+y_i u_j)~, \qquad H(y) = \prod\limits_{i,j} \frac1{1-y_i u_j}~. 
\end{equation}
Because these two generating functions contain the same information, we shall work with $E(y)$ in what follows. 

Expanding $E(y)$ as a formal power series in the parameters $y$, one obtains
\begin{equation}
\label{WL:EH2}
	E(y) = 1 + \sum\limits_{\lambda} e_\lambda(y) m_\lambda(u)~,
\end{equation}
where $\sum_{\lambda}$ denotes the sum over all \emph{non-empty} partitions.\footnote{The unity term corresponding to the empty partition has been separated in order to avoid possible ambiguities, which are present in \cite{ Muck:2019hnz}.} One may call $m_\lambda(u)$ a monomial representation of the Wilson loop. A Wilson loop in some irreducible representation of U$(N)$ is given by a Schur function 
\begin{equation}
\label{WL:Tr.U}
	\Tr_\lambda(U) = s_\lambda(u)~.
\end{equation}
Using the Cauchy identity \cite{Macdonald:1995} in \eqref{WL:EH2} one has, 
\begin{equation}
\label{WL:Cauchy}
	E(y) = 1 + \sum\limits_\lambda  s_\lambda(y) s_{\lambda^\dagger}(u)~.
\end{equation}

Products of multiply wound Wilson loops are given by power-sum functions. More precisely, given a composition $K$, we have
\begin{equation}
\label{WL:prod.loops}
	\prod\limits_{i=1}^{l(K)} \Tr U^{k_i}
	= \prod\limits_{i=1}^{l(K)} p_{k_i}(u) = p_K(u)~.
\end{equation}
Recall that $p_K(u)=p_\lambda(u)$, with $\lambda$ the partition that is associated with $K$, because the traces commute in the product. 
In terms of the power-sum basis, \eqref{WL:EH2} reads
\begin{equation}
\label{WL:power.sum}
	E(y) = 1 + \sum\limits_\lambda \frac{\varepsilon_\lambda}{z_\lambda} 
	p_\lambda(y) p_{\lambda}(u)~,
\end{equation}
where $\varepsilon_\lambda$ is a shorthand for 
\begin{equation}
\label{WL:eps.shorthand}
	\varepsilon_\lambda = (-1)^{|\lambda|-l(\lambda)}~.
\end{equation}
Wilson loop expectation values $\vev{.}$ are now encoded in $\vev{E(y)}$, and connected correlators $\vvev{.}$ are defined in terms of $\ln \vev{E(y)}$. For example, expanding $\ln \vev{E(y)}$ in the power-sum basis,
\begin{equation}
\label{WL:connected.correlators}
	\ln \vev{E(y)} = \sum\limits_\lambda 
	\frac{\varepsilon_\lambda}{z_\lambda}
 	p_\lambda(y) \vvev{p_{\lambda}(u)}~,
\end{equation}
yields the connected correlators of multiply wound Wilson loops. Note that the unity term is absent in \eqref{WL:connected.correlators}.

The above argument can be extended to two or more different Wilson loops.\footnote{By \emph{different} we mean following different contours.} Let us consider two of them, with gauge holonomies $U$ and $\bar{U}$, respectively. We now need two alphabets of parameters, $y$ and $x$, and two generating functions, $E(y)$ and $\bar{E}(x)$. Correlators between the two Wilson loops are encoded in the generating function $\vev{E(y)\bar{E}(x)}$. Moreover, taking the logarithm defines the connected correlators. 
For example,
\begin{equation}
\label{WL:connected.correlators.prod}
	\ln \vev{E(y)\bar{E}(x)} = 
	\ln \vev{E(y)} + \ln\vev{\bar{E}(x)} 
	+ \sum\limits_{\lambda,\nu}
	\frac{\varepsilon_\lambda\varepsilon_\nu}{z_\lambda z_\nu} 
	p_\lambda(y) p_\nu(x) \vvev{p_{\lambda}(u) p_{\nu}(\bar{u})}~.
\end{equation}
The first two terms on the right hand side arise from the unity term in \eqref{WL:power.sum}, and we recall that the sums include only non-empty partitions. Specializing to $x = 0$ reduces \eqref{WL:connected.correlators.prod} to \eqref{WL:connected.correlators}. Moreover, one may consider the special case of equal Wilson loops, $U=\bar{U}$, which implies $\bar{E}(x) = E(x)$. In this case, we simply have $E(y)E(x) = E(y\oplus x)$, where $\oplus$ is the operation of alphabet addition \cite{Lascoux}. Thus, one would not get any new information from the product.

%% file: N4corr.tex
\section{Wilson loop generating functions in $\mathcal{N}=4$ SYM theory}
\label{N4}

\subsection{Review of the single loop case}
\label{N4:review}
In this subsection, we briefly review the results of \cite{Muck:2019hnz}.
In $\mathcal{N}=4$ SYM theory, $\frac12$-BPS circular Wilson loops can be mapped by localization to a Gaussian matrix model \cite{Erickson:2000af,Drukker:2000rr,Akemann:2001st,Hartnoll:2006is,Fiol:2013hna}. Our conventions for the matrix model are
\begin{equation}
\label{N4:MM.def}
	\vev{f(X)}_{\mathrm{mm}} = \int \mathcal{D} X\, f(X) \e{-\frac12 X^2}~, 
	 \qquad  \vev{1}_{\mathrm{mm}} = 1~.
\end{equation}
We shall consider only the case of U$(N)$ as gauge group, in which the matrix integral is over hermitian matrices $X$.\footnote{For SU$(N)$, the matrices must be hermitian and traceless.}

In terms of the Gaussian matrix model, the Wilson loop generating function $\vev{E(y)}$ defined in \eqref{WL:EH} is given by 
\begin{equation}
\label{N4:vevEmm}
	\vev{E(y)} 
	= \vev{\prod\limits_i \det \left( 1 + y_i \e{gX}\right)}_{\mathrm{mm}} 
	= \vev{\det \left[\sum\limits_{k=0}^\infty e_k(y) 
		\e{kgX} \right]}_{\mathrm{mm}} ~,
\end{equation} 
where $g=\frac12 g_{\mathrm{YM}}$. The 't~Hooft coupling is then $\lambda = 4N g^2$. The matrix model integral can be done with standard techniques \cite{Mehta:1981xt}, and the result is \cite{CanazasGaray:2018cpk}
\begin{equation}
\label{N4:vevE}
	\vev{E(y)} = \det \left[ \sum\limits_{k=0}^\infty e_k(y) A_k \right]~,
\end{equation}
where $A_k$ represents the $N \times N$ matrix \cite{Okuyama:2018aij, CanazasGaray:2018cpk}
\begin{equation}
\label{N4:A.expl}
	(A_k)_{m,n} = (A_k)_{n,m}= \sqrt{\frac{n!}{m!}} \e{\frac12 k^2 g^2} \left(k g \right)^{m-n}
		\Laguerre^{m-n}_n\left(-k^2 g^2\right)~.
\end{equation}
$\Laguerre^{m-n}_n(x)$ denotes an associated Lagurerre polynomial, and $m,n=0,\ldots, N-1$. It is important to notice that the integer $k$ only appears in the product $kg$.

Taking the logarithm of \eqref{N4:vevE} yields
\begin{equation}
\label{N4:W.prime.sol}
	\ln \vev{E(y)} = \sum\limits_\lambda e_\lambda(y)  \frac{(-1)^{l(\lambda)-1}}{z_\lambda} [l(\lambda)-1]!
	\left( \prod\limits_i \lambda_i\right) \Tr\left[ A_{(\lambda_1} A_{\lambda_2} \cdots A_{\lambda_{l(\lambda)})} \right]~, 
\end{equation}
which tells us, from \eqref{WL:EH2}, that the traces of the symmetrized matrix products\footnote{Symmetrization includes a normalization factor $1/l(\lambda)!$.} are related to the connected Wilson loop expectation value in the monomial basis, 
\begin{equation}
\label{N4:M.conn}
	\Tr\left[A_{(\lambda)}\right] \equiv 
	\Tr\left[ A_{(\lambda_1} A_{\lambda_2} \cdots A_{\lambda_{l(\lambda)})} \right] = \frac{(-1)^{l(\lambda)-1}}{[l(\lambda)-1]!} 
	\frac{z_\lambda}{\prod_i \lambda_i} \vvev{m_\lambda(u)}~.
\end{equation}
Using purely combinatorial relations, the monomials can be translated into any other basis. For the power-sum basis, which represents the correlators of multiply wound loops, the relation is \cite{Lascoux} 
\begin{equation}
\label{N4:M.conn.2}
	\Tr \left[ A_{(\lambda)} \right] = \frac{(-1)^{l(\lambda)-1}}{[l(\lambda)-1]!} 
	\sum\limits_{\nu\in\mathcal{P}([l(\lambda)])} \mathcal{M}(\nu)\, 
	\vvev{p_{\lambda_\nu}(u)}~.
\end{equation}
In \eqref{N4:M.conn.2}, the sum is over all \emph{set partitions}, $\nu$, of $[l(\lambda)]=(1,2,\ldots,l(\lambda))$, $\mathcal{M}(\nu)$ denotes the M\"obius function 
\begin{equation}
\label{N4:Mobius}
	\mathcal{M}(\nu) = \prod\limits_i (-1)^{l(\nu_i)-1} [l(\nu_i)-1]!~,
\end{equation} 
and the composition $\lambda_\nu$ was defined in \eqref{Intro:K.nu.def}\footnote{One needs to substitute $\lambda$ in place of $K$ in \eqref{Intro:K.nu.def}}.
In \cite{Muck:2019hnz}, the inverse of \eqref{N4:M.conn.2} was given without proof, but we refrain from reviewing it here, because a more general relation will be derived below.

In the above equations, all $\lambda_i$ are \emph{strictly positive}, because $\lambda$ always denotes a proper partition. This implies that all of the loops in the correlator \eqref{WL:prod.loops} have the same orientation. For unitary groups, inverting the orientation of the loop maps each representation to its complex conjugate. In the above calculation, this amounts to swapping the sign of the gauge coupling, $g$. Therefore, denoting the generating function of the Wilson loops in the complex conjugate representations by $\bar{E}(y)$, one has
\begin{equation}
\label{N4:vevE.cc}
	\vev{\bar{E}(y)} = \vev{\prod\limits_i \det \left( 1 + y_i \e{-gX}\right)}_{\mathrm{mm}}
	= \det \left[ \sum\limits_{k=0}^\infty e_k(y) A_{-k} \right]~.
\end{equation}
As is obvious from \eqref{N4:A.expl}, we have $A_{-k}\neq A_k$, the difference arising from the term $(kg)^{m-n}$. However, because this term cancels in matrix products, it holds that $\Tr A_{-\lambda} = \Tr A_\lambda$, where $-\lambda = (-\lambda_1,- \lambda_2, \ldots, - \lambda_{l(\lambda)})$. In turn, together with \eqref{N4:W.prime.sol}, this implies $\vev{\bar{E}(y)} = \vev{E(y)}$. This result is of course expected, because the choice of the common orientation of all the loops is irrelevant.

In order to discuss correlators of loops with opposite orientations, we need to consider the case of two different loops discussed at the end of subsection~\ref{WLcorr:genfunc}. This is what we will do next.

\subsection{Oppositely wound loops}
\label{N4:loops}
In this subsection, our aim is to generalize the results reviewed in the previous subsection to products of multiply would Wilson loops with arbitrary orientation. The loops are still spacially overlapping to ensure that the configuration remains $\frac12$-BPS, so that it can be mapped to the Gaussian matrix model. This is the general case considered by Okuyama \cite{Okuyama:2018aij}. Specifically, we are interested in correlators of the form
\begin{equation}
\label{N4:prod.WL}
	\mathcal{W}^{(k_1,k_2,\ldots,k_h)} = \vev{\prod_{i=1}^h \Tr U^{k_i}}~,
\end{equation}
where $k_1,k_2,\ldots,k_h$ are \emph{non-zero} integers.\footnote{Zeros lead to trivial modifications, because $\Tr U^0=N$.}
The results of the previous subsection can be applied to \eqref{N4:prod.WL}, if either all $k_i$ are positive, or all $k_i$ are negative. These two cases are equivalent, because $U^{-k} = (U^{-1})^k$, but we have seen that $\vev{E(y)}=\vev{\bar{E}(y)}$ holds for representations that are complex conjugates of each other. In the general case, we can collect the positive and negative integers into two sets using the commutativity of the traces and rewrite \eqref{N4:prod.WL} as 
\begin{equation}
\label{N4:prod.WL.2}
	\mathcal{W}^{(\lambda_1,\lambda_2,\ldots,-\nu_1,-\nu_2,\ldots)} =
	\vev{ p_\lambda(u) p_\nu(\bar{u}) }~.
\end{equation}
These correlators belong to the generic two-loop case discussed at the end of subsection~\ref{WLcorr:genfunc}. 

Consider the two-loop generating function $\vev{E(y)\bar{E}(x)}$, where $y$ and $x$ are two independent alphabets of parameters. The matrix model expression of this generating function is   
\begin{align}
\notag
	\vev{E(y)\bar{E}(x)} &= \vev{\prod\limits_{i,j} \det \left[
	\left( 1 + y_i \e{gX}\right)\left( 1 + x_j \e{-gX}\right) \right]
	}_{\mathrm{mm}} \\
\label{N4:vevEbarE}
	&= \det \left[ \sum\limits_{k,l=0}^\infty e_k(y) e_l(x) A_{k-l}
	 \right]~,
\end{align} 
with the $N\times N$ matrices $A_k$ again given in \eqref{N4:A.expl}.

There are different ways to proceed from here. The first way is to rewrite \eqref{N4:vevEbarE} as 
\begin{align}
\notag
	\vev{E(y)\bar{E}(x)} &= 
	\det \left[ 1 
	+ \sum\limits_{k=1}^\infty e_k(y) A_k
	+ \sum\limits_{l=1}^\infty e_l(x) A_{-l}
	+ \sum\limits_{k,l=1}^\infty e_k(y) e_l(x) A_{k-l} \right]\\
\notag 
	&= \det \left\{ \left[ 1 + \sum\limits_{k=1}^\infty e_k(y) A_k \right] 
	\left[ 1 + \sum\limits_{l=1}^\infty e_l(x) A_{-l} \right] +  
	\sum\limits_{k,l=1}^\infty e_k(y) e_l(x) 
	\left(A_{k-l} - A_k A_{-l}\right) \right\}\\
\label{N4:EbarE.1}
	&= \vev{E(y)} \vev{\bar{E}(x)} 
	\det \Bigg\{ 1 + 
	\left[ 1 + \sum\limits_{l=1}^\infty e_l(x) A_{-l} \right]^{-1} 
	\left[1+ \sum\limits_{k=1}^\infty e_k(y) A_k \right]^{-1} \\
\notag &\quad \times
	\sum\limits_{k,l=1}^\infty e_k(y) e_l(x) (A_{k-l} - A_k A_{-l}) \Bigg\}~.
\end{align}
One could now take the logarithm in \eqref{N4:EbarE.1}, which would yield the generating function of the connected correlators \eqref{WL:connected.correlators.prod}. While this would reproduce nicely the first two terms on the right hand side of \eqref{WL:connected.correlators.prod}, expanding the remaining term in $y$ and $x$ would seem dreadfully complicated, because of the matrix inverses. Let alone the conversion to the power-sum basis.

Another way of manipulating \eqref{N4:vevEbarE} is to reorder the double sum and collect the terms with equal $A_k$. This yields
\begin{align}
\notag
	\vev{E(y)\bar{E}(x)} &= 
	\det \left\{ \alpha_0(y,x) 
	+ \sum\limits_{k=1}^\infty \left[ \alpha_k (y,x) A_k 
	+ \alpha_k(x,y) A_{-k} \right] \right\}\\
	&= \alpha_0(y,x)^N \det \left\{ 1 + \sum\limits_{k=1}^\infty \left[
	\frac{\alpha_k(y,x)}{\alpha_0(y,x)}	A_k 
	+ \frac{\alpha_k(x,y)}{\alpha_0(y,x)} A_{-k} \right] \right\}~,
\end{align}
where the functions $\alpha_k(y,x)$ are defined by
\begin{equation}
\label{N4:alpha.def}
	\alpha_k(y,x) = \sum\limits_{l=0}^\infty e_{k+l}(y) e_l(x)~.
\end{equation}
This time, after taking the logarithm, the expansion in powers of $\alpha_k/\alpha_0$ is straightforward, but the conversion into the power sum basis still seems dreadful. Therefore, we shall proceed differently and follow Okuyama \cite{Okuyama:2018aij}. 

Okuyama considered the generating function of multiply wound Wilson loops,
\begin{equation}
\label{N4:gen.func.Oku}
	\mathcal{G}^{(k_1,k_2,\ldots,k_h)}(y) = 
	\vev{\prod\limits_{i=1}^h \det
	\left( 1 +y_i \e{k_i g X} \right)}_{\mathrm{mm}}~,
\end{equation}
where $y$ denotes the finite alphabet of parameters $y=(y_1, y_2, \ldots, y_h)$. Because of 
\[ \det \left( 1 +y_i \e{k_i g X} \right) = 1 + y_i \Tr \e{k_i g X} + \Order(y_i^2)~,\]
the correlator \eqref{N4:prod.WL} is the coefficient of the maximum-rank elementary polynomial $e_h(y)$ in the Taylor expansion of $\mathcal{G}^{(k_1,k_2,\ldots,k_h)}(y)$. Similarly, the connected correlator is the coefficient of $e_h(y)$ in the Taylor  expansion of $\ln \mathcal{G}^{(k_1,k_2,\ldots,k_h)}(y)$. 

Evaluating the matrix model expectation value in \eqref{N4:gen.func.Oku} results in
\begin{equation}
\label{N4:gen.func.Oku.mm}
	\mathcal{G}^{(k_1,k_2,\ldots,k_h)}(y) = \det \left( 1+ 
	\sum_{\emptyset \neq \mu\subseteq[h]} 
	y_\mu A_{k_\mu} \right)~,
\end{equation}
where the sum runs over all \emph{non-empty} subsets of $[h]$, and we defined
\begin{equation}
\label{N4:kmu.def}
	y_\mu = \prod\limits_{i \in \mu} y_{i}~, \qquad 
	k_\mu = \sum\limits_{i\in \mu} k_i~.
\end{equation}
Then, taking the logarithm in \eqref{N4:gen.func.Oku.mm} yields
\begin{equation}
\label{N4:gen.func.Oku.log}
	\ln	\mathcal{G}^{(k_1,k_2,\ldots,k_h)}(y) = \sum\limits_{c=1}^\infty		\frac{(-1)^{c-1}}{c} \Tr \left( \prod\limits_{i=1}^c
	\sum_{\emptyset \neq \mu_i\subseteq[h]} y_{\mu_i} 
	A_{k_{\mu_i}} \right)~.
\end{equation}
In \eqref{N4:gen.func.Oku.log}, the terms containing the maximum-rank elementary polynomial $e_h(y)$ are precisely those in which $(\mu_1, \mu_2, \ldots, \mu_c)$ constitutes a \emph{set partition} of $[h]$ (for the definition of a set partition, see section~\ref{WLcorr:review}), such that each parameter $y_i$ appears exactly once in the product. Therefore,
\begin{equation}
\label{N4:ccorr}
	\mathcal{W}^{(k_1,k_2,\ldots,k_h)}_\text{conn} \equiv \vvev{\prod_{i=1}^h \Tr U^{k_i}} = 
	\sum\limits_{\nu\in \mathcal{P}([h])} \frac{(-1)^{l(\nu)-1}}{l(\nu)}  
	\Tr A_{K_\nu}~,
\end{equation}
where the sum is over all set partitions of $[h]$. In analogy with \eqref{Intro:K.nu.def}, $K_\nu$ denotes the set\footnote{Here, we cannot call $K_\nu$ a composition, because the integers $k_i$ are not necessarily positive.}
\begin{equation}
\label{N4:k.nu.def}
	K_\nu = \left( \sum_{i\in \nu_1} k_i, \sum_{i\in \nu_2} k_i, 
	\ldots, \sum_{i\in \nu_{l(\nu)}} k_i \right)~,  
\end{equation}
and we introduced
\begin{equation}
\label{N4:Ak.def}
	A_{K_\nu} = \prod\limits_{i=1}^{l(\nu)} A_{(K_\nu)_i}~.
\end{equation}
Eqn.\ \eqref{N4:ccorr} is our final result of this subsection, which generalizes the expressions given by Okuyama \cite{Okuyama:2018aij} to arbitrary $h$. It is also equivalent to formula (4.18) of \cite{Muck:2019hnz}, which it generalizes to arbitrary integers $k_i$.\footnote{Formula (4.18) of \cite{Muck:2019hnz} is expressed in terms of a partition $\lambda$ and contains explicit symmetrizations both over the $k_i$s and over the matrix products. To establish the equivalence with \eqref{N4:ccorr}, one can use the unique association of a partition $\lambda$ with a given set partition, as explained in  \cite{Muck:2019hnz}. The factor $|\mathcal{P}_\lambda|$ stems from the multiplicity of set partitions associated to the same partition $\lambda$. The other factor $l(\lambda)!$ is the normalization factor in the symmetrized product of matrices.}

To end this section, let us consider the special (trivial) case when at least one of the integers $k_i$ is zero. We simply have
\begin{equation}
\label{N4:h0}
	\vev{\Tr U^{0}} = \Tr\left( A_{0} \right) = N~,
\end{equation}
and 
\begin{equation}
\label{N4:h20}
	\vvev{\Tr U^{0} \Tr U^{k_2}\ldots\Tr U^{k_h}} =0 
	\qquad (h\geq2)~.
\end{equation}
Eqn.\ \eqref{N4:h20} is a consequence of $\vev{1 O} = \vev{1}\vev{O}$ for any operator $O$, which means that the connected part of the correlator is trivial. To see this explicitly in \eqref{N4:ccorr}, take $k_1=0$ and separate the set partitions into two groups. In the first group, the integer $1$ sits alone in a set, in the second group not. Consider first a set partition $\nu$ with $l(\nu)=c$ parts, which belongs to the first group, \ie in which one part is $\nu_i=\{1\}$, and the remainder $\nu'= \nu/\nu_i$ is a set partition of $\{2,3,\ldots,h\}$ with $l(\nu')=c-1$ parts. There are $c$ equivalent choices for $i$, so that these set partitions contribute 
\[ c \frac{(-1)^{c-1}}{c} \Tr A_{k_\nu} = (-1)^{c-1} \Tr A_{k_{\nu'}} \]
to \eqref{N4:ccorr}. Compare this to the contribution of the set partitions $\nu''$ of length $c-1$ that belong to the second group. These set partitions are obtained by adding the number $1$ to one of the $c-1$ parts of $\nu'$ defined above. Their contribution to \eqref{N4:ccorr} is 
\[ (c-1) \frac{(-1)^{c-2}}{c-1} \Tr A_{k_{\nu''}} 
= - (-1)^{c-1} \Tr A_{k_{\nu'}}~. \]
Thus, the two contributions cancel, which proves \eqref{N4:h20} after iterating through all $c$.

%% file: exact.tex
\section{Exact results from the matrix model}
\label{exact}

\subsection{Matrix model results from harmonic oscillator}

In this section, we will review and elaborate on exact results that can be obtained from the Gaussian matrix model \eqref{N4:MM.def}. Our analysis will be based on the very elegant treatment of Okuyama \cite{Okuyama:2018yep} and exploits the intricate relation between the hermitian matrix model and the simple harmonic oscillator quantum mechanics. Mathematically, this relation appears, because the Vandermonde determinant, which is introduced in the matrix integral by the matrix diagonalization, is most conveniently expressed in terms of Hermite polynomials \cite{Mehta:1981xt}, which represent the eigenfunctions of the harmonic oscillator \cite{CanazasGaray:2018cpk, Okuyama:2018aij}. Therefore, the matrices $A_k$ given in \eqref{N4:A.expl} are nothing but the matrix elements \cite{Okuyama:2018yep}
\begin{equation}
\label{ex:A.matrix}
	(A_k)_{ij} = \bra{i} \e{kg(a+a^\dagger)} \ket{j}~,
\end{equation}
where $a$ and $a^\dagger$ are the oscillator lowering and raising operators satisfying 
\begin{equation}
\label{ex:osc.op}
	[a,a^\dagger ] = 1~,
\end{equation}
and the states $\ket{i}$ are the normalized eigenstates of the number operator,
\begin{equation}
\label{ex:states}
	\ket{i} = \frac{(a^\dagger)^i}{\sqrt{i!}} \ket{0}~.
\end{equation} 

Before going on, let us slightly change notation by introducing 
\begin{equation}
\label{ex:z.def}
	z=kg =	k \sqrt{\frac{\lambda}{4N}}~,
\end{equation}
which can be treated as a continuous (real or complex) variable. Because $k$ appears in $A_k$ only within $z$, we will also let $A_z \equiv A_k$, so that \eqref{ex:A.matrix} reads
\begin{equation}
\label{ex:A.matrix.z}
	(A_z)_{ij} = \bra{i} \e{z(a+a^\dagger)} \ket{j}~.
\end{equation}
Whereas the harmonic oscillator eigenstates are given by $i,j=0,1,\ldots, \infty$, the matrix model involves only the elements $i,j =0,1,\ldots,N-1$.\footnote{The relation between the matrix $A_z$ and the algebra of the truncated harmonic oscillator was made explicit in \cite{CanazasGaray:2018cpk}. The truncated harmonic oscillator is defined by $N\times N$ matrix lowering and raising operators $b$ and $b^\dagger$ satisfying $[b,b^\dagger] = 1 -N P_{N-1}$, where $P_{N-1}$ is the projector onto the highest eigenstate. This is required by the fact that the trace of any commutator must vanish in a finite-dimensional system, in contrast to the infinite-dimensional system of the standard harmonic oscillator. Nevertheless, the number operator $n_b = b^\dagger b$ retains the standard commutators $[n_b,b] = -b$ and $[n_b,b^\dagger] = b^\dagger$.}
The clever insight of Okuyama \cite{Okuyama:2018yep} is that one can work in the infinite-dimensional Hilbert space, if one truncates any sum over the eigenstates  by inserting the projector
\begin{equation}
\label{ex:project}
	P = \sum\limits_{i=0}^{N-1} \ket{i}\bra{i}~.
\end{equation}
The elegance of this approach can already be seen in the calculation of the matrix element \eqref{ex:A.matrix.z}, which we wish to report here from \cite{Okuyama:2018yep}. One starts with rewriting \eqref{ex:A.matrix.z} as
\begin{align}
\notag
	(A_z)_{ij} &= \e{-\frac{z^2}2} \bra{i} \e{za}\e{za^\dagger} \ket{j} 
	= \frac{\e{-\frac{z^2}2}}{\sqrt{i!j!}}  \bra{0} a^i\e{za}\e{za^\dagger}
	(a^\dagger)^j \ket{0}\\
\notag
	&= \frac{\e{-\frac{z^2}2}}{\sqrt{i!j!}} \partial^i_s \partial^j_t 
	\bra{0} \e{(z+s)a}\e{(z+t)a^\dagger} \ket{0} |_{s=t=0}\\
\notag
	&= \frac{\e{-\frac{z^2}2}}{\sqrt{i!j!}} \partial^i_s \partial^j_t 
	\e{(z+s)(z+t)} |_{s=t=0}\\
\label{ex:A.calc}
	&= \frac{\e{\frac{z^2}2}}{\sqrt{i!j!}} \partial^i_s 
	(z+s)^j \e{zs} |_{s=0}~.
\end{align}
After introducing $w=\frac{z}{s}$, the generating function of the Laguerre polynomials \cite{Gradshteyn} can be recognized, so that \eqref{ex:A.calc} becomes
\begin{align}
\notag
	(A_z)_{ij} &= \frac{\e{\frac{z^2}2}}{\sqrt{i!j!}} z^{j-i} \partial^i_w
	\sum\limits_{n=0}^\infty \Laguerre_n^{j-n} (-z^2) w^n |_{w=0}\\
\label{ex:A.result}
	&= \sqrt{\frac{i!}{j!}} \e{\frac{z^2}2} z^{j-i} L_i^{j-i}(-z^2)~,
\end{align}
reproducing \eqref{ex:A.matrix.z}.

\subsection{One-point functions}
Although the calculation of the one-point function
\begin{equation}
\label{ex:1pt.def}
	\mathcal{W}^{(k)} \equiv \vev{\Tr U^k} = \Tr A_z
\end{equation}
is very easy by tracing over the matrix \eqref{ex:A.result}, it is instructive to do the calculation in the infinite-dimensional system \cite{Okuyama:2018yep}. One starts with 
\begin{equation}
\label{ex:1pt.new}
	\mathcal{W}^{(k)} = \Tr_\infty \left(\e{z(a+a^\dagger)} P\right)~,
\end{equation}
where $\Tr_\infty$ denotes the trace in the infinite-dimensional Hilbert space. Then, one exploits the relations
\begin{equation}
\label{ex:z.exp.rels}
	z \e{z(a+a^\dagger)} = [a, \e{z(a+a^\dagger)}] 
	= [\e{z(a+a^\dagger)}, a^\dagger]~,
\end{equation}
the cyclic property of the trace, as well as the commutators 
\begin{equation}
\label{ex:a.P.comm}
	[a,P] = -\sqrt{N} \ket{N-1}\bra{N}~,\qquad 
	[a^\dagger,P] = \sqrt{N} \ket{N}\bra{N-1}~.
\end{equation}
Thus, one can write
\begin{align}
\notag
	z \mathcal{W}^{(k)} &= \Tr_\infty \left( z \e{z(a+a^\dagger)} P\right) 
	=  \Tr_\infty \left( [a,\e{z(a+a^\dagger)}] P\right)
	=  \Tr_\infty \left( \e{z(a+a^\dagger)} [P,a] \right)\\
\notag
	&= \sqrt{N} \bra{N} \e{z(a+a^\dagger)} \ket{N-1} 
	= \sqrt{N} A_{N,N-1} = \sqrt{N} A_{N-1,N}\\
\label{ex:Tr.A.result}
	&= z \e{\frac{z^2}2} \Laguerre^1_{N-1}(-z^2)~,
\end{align}
which reproduces the known result
\begin{equation}
\label{ex:Tr.A.result.2}
	\mathcal{W}^{(k)} = \e{\frac{z^2}2} \Laguerre^1_{N-1}(-z^2)~.
\end{equation}

\subsection{Two-point functions}

Let us extend the procedure of the previous subsection to the two-point function $\Tr(A_{z_1}A_{z_2})$. A similar calculation yields \cite{Okuyama:2018yep}
\begin{equation}
\label{ex:2pt.calc}
	(z_1+z_2) \Tr(A_{z_1}A_{z_2}) = \sqrt{N} \bra{N}  
	\e{z_1(a+a^\dagger)} P \e{z_2(a+a^\dagger)} \ket{N-1} 
	+ (z_1 \leftrightarrow z_2)~. 
\end{equation}
Furthermore, one can write (abbreviating $\partial_n = \partial_{z_n}$ for $n=1,2$)
\begin{align}
\notag
	(z_1+z_2)(\partial_1-\partial_2) \Tr(A_{z_1}A_{z_2}) 
	&= \sqrt{N} \bra{N} \e{z_1(a+a^\dagger)} [a+a^\dagger, P] \e{z_2(a+a^\dagger)}
	\ket{N-1} - (z_1 \leftrightarrow z_2) \\
\notag 
	&= N \left[ (A_{z_1})_{N,N}(A_{z_2})_{N-1,N-1}
		- (A_{z_2})_{N,N}(A_{z_1})_{N-1,N-1} \right]\\
\label{ex:2pt.Okuyama}
	&=  N \e{\frac{z_1^2+z_2^2}2} \left[ 
	\Laguerre_N^0(-z_1^2) \Laguerre_{N-1}^0(-z_2^2)
	-\Laguerre_N^0(-z_2^2) \Laguerre_{N-1}^0(-z_1^2) \right]~.
\end{align}
This is where the calculation stops in \cite{Okuyama:2018yep}, and we will take it from there. First, using the sum \cite[8.974.1]{Gradshteyn}, \eqref{ex:2pt.Okuyama} can be rewritten as 
\begin{equation}
	(z_1+z_2)(\partial_1-\partial_2) \Tr(A_{z_1}A_{z_2}) 
	= (z_1^2-z_2^2) \e{\frac{z_1^2+z_2^2}2} 
	\sum\limits_{m=0}^{N-1} \Laguerre^0_m(-z_1^2)\Laguerre^0_m(-z_2^2)~,
\end{equation}
which gives 
\begin{equation}
\label{exact:2pt.inter}
	(\partial_1-\partial_2) \Tr(A_{z_1}A_{z_2}) 
	= (z_1-z_2) \e{\frac{z_1^2+z_2^2}2} 
	\sum\limits_{m=0}^{N-1} \Laguerre^0_m(-z_1^2)\Laguerre^0_m(-z_2^2)~.
\end{equation}
Then, using \cite[8.976.4]{Gradshteyn}, \eqref{exact:2pt.inter} is equal to 
\begin{equation}
	(\partial_1-\partial_2) \Tr(A_{z_1}A_{z_2}) 
	= (z_1-z_2) \e{\frac{z_1^2+z_2^2}2} 
	\sum\limits_{m=0}^{N-1} \sum\limits_{k=0}^m \Laguerre^{2k}_{m-k}(-z_1^2-z_2^2)
	\frac{(z_1z_2)^{2k}}{(k!)^2}~.
\end{equation}
After reordering the summations, one obtains
\begin{align}
\notag 
	(\partial_1-\partial_2) \Tr(A_{z_1}A_{z_2}) 
	&= (z_1-z_2) \e{\frac{z_1^2+z_2^2}2} \sum\limits_{k=0}^{N-1}
	\sum\limits_{m=0}^{N-1-k}\Laguerre^{2k}_{m}(-z_1^2-z_2^2)
	\frac{(z_1z_2)^{2k}}{(k!)^2}\\
\label{ex:delAA}
	&= (z_1-z_2) \e{\frac{z_1^2+z_2^2}2} \sum\limits_{k=0}^{N-1}
	\Laguerre^{2k+1}_{N-1-k}(-z_1^2-z_2^2)
	\frac{(z_1z_2)^{2k}}{(k!)^2}~.
\end{align}
We will further manipulate this expression and integrate it in section~\ref{tp}.  

%% file: fundamental.tex
\section{One-point functions}
\label{onept}

In this section, we revisit the one-point functions $\mathcal{W}^{(k)} = \vev{\Tr U^k}$, with $k$ of order unity (as opposed to order $N$ or $\sqrt{N}$, for example). Without loss of generality, we can set $k=1$, keeping in mind that $k$ appears in the exact result  \eqref{ex:Tr.A.result.2} only in the combination $z^2=k^2\frac{\lambda}{4N}$. The general case can be recovered from the case $k=1$ by scaling $\lambda\to k^2 \lambda$. The $1/N$ expansion of $\mathcal{W}^{(1)}$ was obtained by Drukker and Gross \cite{Drukker:2000rr}. We will review their solution and provide a very simple check of it. The solution contains certain numerical coefficients, which are defined through a recursive procedure. Then, we will present an explicit construction, which results in a direct combinatorial formula for these coefficients.

\subsection{Drukker and Gross' series in $1/N$}
\label{onept:DG}

The expansion into a series in $1/N$ of the exact result \eqref{ex:Tr.A.result.2},
\begin{equation}
\label{fund:W1}
	\mathcal{W}^{(1)}
	= \e{\frac{\lambda}{8N}} \Laguerre^{1}_{N-1}
		\left(-\frac{\lambda}{4N}\right)~,
\end{equation}
has the form \cite{Drukker:2000rr}
\begin{equation}
\label{fund:W1.expand}
	\mathcal{W}^{(1)} = \sum\limits_{m=0}^\infty N^{1-2m} W_m~,  
\end{equation}
where the genus-$m$ contributions $W_m$ are given by\footnote{Our coefficients $B(m,k)$ are related to Drukker and Gross' $X^i_k$ by $B(m,k)= X_m^{m-k}$. Morever, we include the term with $m=k=0$ in the sum.}
\begin{equation}
\label{fund:W.m}
	W_m = \sum\limits_{k=0}^m B(m,k) 
	\left(\frac{\sqrt{\lambda}}{2}\right)^{2m+k-1}
		\BesselI[2m+k-1](\sqrt{\lambda})~.
\end{equation}
Here, $\BesselI[\alpha](z)$ are the modified Bessel functions of the first kind, and the coefficients $B(m,k)$ are determined by the recurrence relation (for $m,k>0$) 
\begin{equation}
\label{fund:B.rec}
	(2m+k) B(m,k) = \frac14(2m+k-2) B(m-1,k) + \frac14 B(m-1,k-1)~,
\end{equation}
together with the initial values
\begin{equation}
\label{fund:B.ini}
	B(m,0) = \delta_{m,0}~,\qquad B(0,k) = \delta_{k,0}~.
\end{equation}
\begin{table}[th]
\caption{Some values of $B(m,k)$. \label{fund:B.tab}}
\[
\left(\begin{array}{rrrrrrr}
 & k & \rightarrow \\
m & 1 & 0 & 0 & 0 & 0 & 0 \\
\downarrow& 0 & \frac{1}{12} & 0 & 0 & 0 & 0 \\
&0 & \frac{1}{80} & \frac{1}{288} & 0 & 0 & 0 \\
&0 & \frac{1}{448} & \frac{1}{960} & \frac{1}{10368} & 0 & 0 \\
&0 & \frac{1}{2304} & \frac{71}{268800} & \frac{1}{23040} & \frac{1}{497664} & 0 \\
&0 & \frac{1}{11264} & \frac{31}{483840} & \frac{23}{1612800} & \frac{1}{829440} & \frac{1}{29859840}
\end{array}\right)
\]
\end{table}

Some values of $B(m,k)$ are listed in Table~\ref{fund:B.tab}. Notice that $B(m,k)=0$ for $k>m$. Other particular values are 
\begin{equation}
\label{fund:B.spec}
	B(m,1) = \frac1{4^m(2m+1)}~, \qquad 
	B(m,m) = \frac1{12^m m!}~, \qquad   
	B(m,m-1) = \frac1{5 \cdot 4^m \cdot 3^{m-2}(m-2)!}~.
\end{equation} 

We find it useful to express $W_m$ in terms of generalized hypergeometric series, see \eqref{apphyp:spec.3}. In particular, for $m=0$,
\begin{equation}
\label{fund:W.0.hyp}
	W_0 = \frac2{\sqrt{\lambda}} \BesselI[1](\sqrt{\lambda}) = 
	\genhypF{0}{1}{-;2;\frac{\lambda}{4}}~.
\end{equation}
For $m>0$, we have
\begin{equation}
\label{fund:W.m.hyp}
	W_m = \sum_{k=1}^m \frac{C(m-1,k-1)}{4^m (2m+k)!} \left(\frac{\lambda}{4}\right)^{2m+k-2} \genhypF{0}{1}{-;2m+k;\frac{\lambda}{4}}~, 
\end{equation}
where we have introduced the new coefficients 
\begin{equation}
\label{fund:C.def}
	C(m,k) = 4^{m+1} (2m+k+3) B(m+1,k+1)~.
\end{equation}
The recurrence relation for $C(m,k)$ is easily found from \eqref{fund:B.rec} and reads
\begin{equation}
\label{fund:C.rec}
	C(m,k) =  C(m-1,k) + \frac{C(m-1,k-1)}{(2m+k)}~,
\end{equation}
with the start values $C(0,0)=12 B(1,1) = 1$ and $C(m,-1)=C(-1,k)=0$. Some values are listed in Table~\ref{fund:C.tab}.

\begin{table}[ht]
\caption{Some values of $C(m,k)$. \label{fund:C.tab}}
\[
\left(\begin{array}{rrrrrr}
& k & \rightarrow\\ 
m& 1 & 0 & 0 & 0 & 0 \\
\downarrow& 1 & \frac{1}{3} & 0 & 0 & 0 \\
&1 & \frac{8}{15} & \frac{1}{18} & 0 & 0  \\
&1 & \frac{71}{105} & \frac{11}{90} & \frac{1}{162} & 0 \\
&1 & \frac{248}{315} & \frac{299}{1575} & \frac{7}{405} & \frac{1}{1944}
\end{array}\right)
\]
\end{table}

There is a slightly different form of \eqref{fund:W.m.hyp}, which we wish to derive for later purposes. Let us first use a recurrence relation for the modified Bessel function to write \eqref{fund:W.m} as
\begin{equation}
\label{fund.W.m.hyp.1.5}
	W_m = \sum\limits_{k=0}^m B(m,k) \left[ 
		\left(\frac{\sqrt{\lambda}}{2}\right)^{2m+k-1} 
		\BesselI[2m+k+1](\sqrt{\lambda})
		+(2m+k) \left(\frac{\sqrt{\lambda}}{2}\right)^{2m+k-2} 
		\BesselI[2m+k](\sqrt{\lambda}) \right]~.
\end{equation}
Next, consider the second term in the bracket in \eqref{fund.W.m.hyp.1.5}. For $k=0$, this term does not contribute to the sum ($B(m,0)=0$ for $m>0$), nor would it for  $k=m+1$, so we can safely shift the summation index by one for this term. After this, the two terms can be combined using \eqref{fund:B.rec} to give
\begin{align}
\notag 
  W_m
  	&= 4 \sum\limits_{k=0}^m (2m+k+3) B(m+1,k+1)  
		\left(\frac{\sqrt{\lambda}}{2}\right)^{2m+k-1} 
		\BesselI[2m+k+1](\sqrt{\lambda})\\
\label{fund:W.m.hyp.2}
	&= \sum\limits_{k=0}^m \frac{C(m,k)}{4^m(2m+k+1)!} 
	\left(\frac{\lambda}{4}\right)^{2m+k} 
	\genhypF{0}{1}{-;2m+k+2;\frac{\lambda}4}~.
\end{align}
The expression \eqref{fund:W.m.hyp.2} is valid for all $m\geq 0$, but it has one summand more than \eqref{fund:W.m.hyp} for $m>0$.

To the best of my knowledge, the easiest way to check the series of Drukker and Gross (and find it, as one might say with hindsight) is as follows. Consider the exact solution \eqref{fund:W1}. Laguerre polynomials satisfy the differential equation \cite{Gradshteyn}
\begin{equation}
\label{fund:Lag.deq}
	x \frac{\rmd^2 \Laguerre^\alpha_n(x)}{\rmd x^2} + 
	(\alpha - x +1)\frac{\rmd \Laguerre^\alpha_n(x)}{\rmd x} 
	+n \Laguerre^\alpha_n(x)=0~.
\end{equation}
With the help of \eqref{fund:Lag.deq} it is straightforward to verify that \eqref{fund:W1} satisfies the differential equation
\begin{equation}
\label{fund:W1.deq}
	\left( \partial_\lambda^2 +\frac{2}{\lambda} \partial_\lambda 
	- \frac1{4\lambda} \right) \mathcal{W}^{(1)} = \frac1{64N^2} \mathcal{W}^{(1)}~.
\end{equation}
Knowing that $\mathcal{W}^{(1)}$ has an expansion of the form \eqref{fund:W1.expand}, \eqref{fund:W1.deq} implies a differential recurrence relation for the genus-$m$ contributions $W_m$,
\begin{equation}
\label{fund:Wm.deq}
	\left( \partial_\lambda^2 +\frac{2}{\lambda} \partial_\lambda 
	- \frac1{4\lambda} \right) W_m = \frac1{64} W_{m-1}~.
\end{equation}
The leading term $W_0$ is the unique homogeneous solution of \eqref{fund:Wm.deq} (up to a normalization constant) that can be written as a power series in $\lambda$. Furthermore, one can show that the functions $W_m$ given in \eqref{fund:W.m} [or \eqref{fund:W.m.hyp}] satisfy \eqref{fund:Wm.deq}. In doing so, one must use some recurrence relation of the modified Bessel functions (or the generalized hypergeometric functions), and the recurrence relation \eqref{fund:B.rec} [or \eqref{fund:C.rec}] is essential. \emph{Vice versa}, given $W_{m-1}$ for some $m>0$, the function $W_m$ is the unique particular solution of \eqref{fund:Wm.deq}, if one demands that it be a power series in $\lambda$ and start with $\lambda^{2m}$.

\subsection{Explicit construction}
\label{fund:explicit}

In this subsection, we shall present a new explicit construction of $\mathcal{W}^{(1)}$ as a series in $1/N$. Our immediate aim is to find non-recursive expressions for the coefficients $B(m,k)$ and $C(m,k)$. The calculation will also serve as a blueprint for the analogous calculation in the case of the two-point function, which we consider in the next section.

Let us start with the exact expression \eqref{fund:W1}, which can be rewritten in terms of a Whittaker function as \cite[18.11.2]{NIST}
\begin{equation}
\label{fund:W1.1}
	\mathcal{W}^{(1)}
	= - \frac{4N^2}{\lambda} \operatorname{M}_{N,\frac12}
		\left(-\frac{\lambda}{4N}\right)~.
\end{equation}
This, in turn, allows for the series expansion \cite[13.14.6]{NIST}
\begin{equation}
\label{fund:W1.2}
	\mathcal{W}^{(1)}
	= N \sum\limits_{n=0}^\infty \frac1{n!} \hypF{-n,1-N;2;2}
		\left(\frac{\lambda}{8N}\right)^n~.
\end{equation}
This expression reproduces (A.8) and (A.9) of \cite{Drukker:2000rr}. Because $\hypF{-n,1-N;2;2}$ is a polynomial of degree $n$ in $N$, and because of the identity \cite{Gradshteyn}
\begin{equation}
\label{fund:hypF.ident}
	\hypF{-n,1-N;2;2} = (-1)^n \hypF{-n,1+N;2;2}~,
\end{equation}
we see that it must have the form 
\begin{equation}
\label{fund:hypF.form}
	\hypF{-n,1-N;2;2} = N^n P_{[\frac{n}2]} \left(N^{-2}\right)~, 
\end{equation}
where $P_{[\frac{n}2]}$ is some polynomial of degree $[\frac{n}2]$. This is far from obvious by direct inspection of the hypergeometric series. To get a procedure where this property is evident, we can use either of the identities \cite{Gradshteyn}
\begin{align}
\label{fund:hypF.ident2}
	\hypF{-n,1-N;2;2} &= \frac{(-1)^n(1-N)_n}{(2)_n}\hypF{-n,1+N;N-n;-1} \\
	&= \frac{(1+N)_n}{(2)_n}\hypF{-n,1-N;-N-n;-1}~.
\end{align}
Using the expression on the second line, writing out the hypergeometric series and simplifying the Pochhammer symbols, one finds
\begin{equation}
\label{fund:hypF.sum}
	\hypF{-n,1-N;2;2} = \frac{1}{N(n+1)!} \sum\limits_{l=0}^{n} 
	\begin{pmatrix} n\\ l \end{pmatrix} (N-l)_{n+1}~.
\end{equation}
We remark that this has the form of a Meixner polynomial \cite{NIST} in $N$. To continue, we write the Pochhammer symbol in \eqref{fund:hypF.sum} as
\begin{equation}
	(N-l)_{n+1} = \frac1N (N-l)_{l+1} (N)_{n-l+1}
\end{equation}  
and note that $(N-l)_{l+1}$ and $(N)_{n-l+1}$ are lowering and rising factorials of $N$, respectively. These can be expanded in terms of the (signed and unsigned) Stirling numbers of the first kind,\footnote{We denote by $s(n,k)$ the signed Stirling numbers of the first kind, the unsigned ones being simply $(-1)^{n-k}s(n,k)$.}
\begin{align}
	(N-l)_{l+1} &= \sum\limits_{k=0}^l s(l+1,k+1) N^{k+1}~,\\ 
    (N)_{n-l+1} &= \sum\limits_{k=0}^{n-l} (-1)^{n-l-k} s(n-l+1,k+1) N^{k+1}~.
\end{align}
Therefore, 
\begin{align}
\notag
	(N-l)_{n+1} &=  \sum\limits_{k=0}^l\sum\limits_{p=0}^{n-l} 
	(-1)^{n-l-p} s(l+1,k+1) s(n-l+1,p+1) N^{k+p+1}\\
\label{fund:Poch.expand}
	&= \sum\limits_{m=0}^n N^{1+n-m} 
	\sum\limits_k (-1)^{l+m+k} s(l+1,k+1) s(n-l+1,n-m-k+1)~,
\end{align}
where we have reordered the double summation, and the sum over $k$ is over all values for which the summand is non-vanishing.

After putting everything back into \eqref{fund:W1.2}, one gets
\begin{align}
\notag
	\mathcal{W}^{(1)} &= \sum\limits_{n=0}^\infty \frac1{n!(n+1)!}
		\left(\frac{\lambda}{8N}\right)^n
		\sum\limits_{l=0}^{n} 
	\begin{pmatrix} n\\ l \end{pmatrix} (N-l)_{n+1}\\
\notag
	&= \sum\limits_{n=0}^\infty \frac1{n!(n+1)!}
		\left(\frac{\lambda}{8}\right)^n \sum\limits_{m=0}^n N^{1-m}\\
	&\quad \times
		\sum\limits_{l=0}^{n} \begin{pmatrix} n\\ l \end{pmatrix}
		\sum\limits_k (-1)^{l+m+k} s(l+1,k+1) s(n-l+1,n-m-k+1)~.
\end{align}
Then, reordering the summations over $n$ and $m$ gives
\begin{align}
\label{fund:W1.3}
	\mathcal{W}^{(1)} &= \sum\limits_{m=0}^\infty N^{1-m}
	\sum\limits_{n=0}^\infty \frac1{(n+m)!(n+m+1)!}
		\left(\frac{\lambda}{8}\right)^{n+m} \\
\notag
	&\quad \times
		\sum\limits_{l=0}^{n+m} \begin{pmatrix} n+m\\ l \end{pmatrix}
		\sum\limits_k (-1)^{l+m+k} s(l+1,k+1) s(n+m-l+1,n-k+1)~.
\end{align}
To see that the terms with odd $m$ are absent in \eqref{fund:W1.3}, we can relabel the summation indices on the second line by $k\to n-k$ and $l \to n+m-l$, which returns the summands with parity $(-1)^m$. Therefore, after dropping the terms with odd $m$, we can read off the genus-$m$ contributions to \eqref{fund:W1.expand} as  
\begin{equation}
\label{fund:W.m.3}
	W_m = \sum\limits_{n=0}^\infty 
	\left(\frac{\lambda}{4}\right)^{n+2m} A(n,m)~,
\end{equation}
where the coefficients $A(n,m)$ are given explicitly by 
\begin{align}
\label{fund:A.expl}
	A(n,m) &= \frac{2^{-n-2m}}{(n+2m)!(n+2m+1)!} \sum\limits_{k=0}^n
		\sum\limits_{l=0}^{2m} 
		\begin{pmatrix} n+2m\\ k+l \end{pmatrix}\\
	\notag
		&\quad \times
		(-1)^{l} s(k+l+1,k+1) s(n-k+2m-l+1,n-k+1)~.
\end{align}

We remark that an alternative representation of $A(n,m)$ can be found by a similar calculation that starts with the hypergeometric series $\hypF{-n,1-N;2;2}$. It yields 
\begin{equation}
\label{fund:A.expl.2}
	A(n,m) = \frac1{(n+2m)!} \sum\limits_{l=0}^{2m} 2^{-l} 
	\begin{pmatrix} n+2m\\ l \end{pmatrix}
	\frac{s(n+2m-l+1,n+1)}{(n+2m-l+1)!}~.
\end{equation} 
In this approach, however, the vanishing of the terms with even powers of $N$ is not obvious from the explicit expression and must be checked by other means. Moreover, to establish the equivalence of \eqref{fund:A.expl} and \eqref{fund:A.expl.2}, some convolution formula of the Stirling numbers \cite{Agoh:2010} might be employed. In any case, we have verified using computer algebra \cite{sagemath} that the two formulae give the same values. 

Another remark is that one can use the recurrence relations of the Stirling numbers and the binomial coefficients to show that the coefficients $A(n,m)$ satisfy the recurrence relation 
\begin{equation}
\label{fund:A.rec}
	(n+2m)(n+2m+1) A(n,m) = A(n-1,m) + \frac14 A(n,m-1)~.
\end{equation}  
This is equivalent to the recurrence relation \eqref{fund:C.rec}. Moreover, \eqref{fund:A.rec} implies the differential equation \eqref{fund:Wm.deq} and, in turn, \eqref{fund:W1.deq}.

At this point, we can make contact with the Drukker-Gross series. Taking \eqref{fund:W.m.hyp.2} and substituting the generalized hypergeometric series, one finds 
\begin{equation}
\label{fund:W.m.5}
	W_m = 4^{-m} \sum\limits_{k=0}^m C(m,k) \sum\limits_{l=0}^\infty
	\frac1{l!(2m+k+l+1)!} \left( \frac{\lambda}4 \right)^{2m+k+l}~.
\end{equation}
Because $C(m,k)=0$ for $k>m$, we can extend the sum over $k$ to infinity and reorder the two sums by setting $n=k+l$. This yields
\begin{equation}
\label{fund:W.m.6}
	W_m = 4^{-m} \sum\limits_{n=0}^\infty \left( \frac{\lambda}4 \right)^{2m+n}
	\sum\limits_{k=0}^{\min(m,n)} \frac{C(m,k)}{(n-k)!(2m+n+1)!}~. 
\end{equation}
Confronting this with \eqref{fund:W.m.3}, we can read off
\begin{equation}
\label{fund:AC.rel}
	A(n,m) = \frac1{4^m(2m+n+1)!} \sum\limits_{k=0}^{\min(m,n)} \frac{C(m,k)}{(n-k)!}~.
\end{equation}
The inverse of this relation is 
\begin{equation}
\label{fund:CA.rel}
	C(m,k) = 4^m \sum\limits_{n=0}^{k} \frac{(-1)^{k-n}(2m+n+1)!}{(k-n)!} A(n,m)~,\qquad (k\leq m)~.
\end{equation}
Finally, combining \eqref{fund:CA.rel} with \eqref{fund:A.expl} or \eqref{fund:A.expl.2} yields explicit expressions for the coefficients $C(m,k)$, without the need of a recursion.\footnote{For the sake of a computer algebra implementation, using the recurrence relation is faster.} For example, with \eqref{fund:A.expl.2},
\begin{equation}
\label{fund:C.expl}
	C(m,k) = \sum\limits_{n=0}^k \frac{(-1)^{k-n}}{(k-n)!}
	\sum\limits_{l=0}^{2m} 2^l \begin{pmatrix} 2m+n+1 \\ 2m-l \end{pmatrix}
	\frac{s(n+1+l,n+1)}{(n+l)!}~.	
\end{equation}

%% file: two-point.tex
\section{Connected two-point functions}
\label{tp}

In this section, we will evaluate the $1/N$ expansion of the connected two-point functions\footnote{We recall our definitions $z=kg$ and $A_z\equiv A_k$.}
\begin{equation}
\label{tp:wconn}
	\mathcal{W}^{(k_1,k_2)}_\text{conn} = \vvev{\Tr U^{k_1}\Tr U^{k_2}} = \Tr A_{z_1+z_2} - \Tr (A_{z_1} A_{z_2})~.
\end{equation}
In subsection~\ref{tp:explicit}, we will perform an explicit calculation along the same line as we did for the one-point function in subsection~\ref{fund:explicit}. This will result in a series in $1/N^2$ with coefficients that are series in $\lambda$ and functions of $k_1$ and $k_2$. In subsection~\ref{tp:leading}, the leading term will be compared to known expressions from the literature. In subsection~\ref{tp:recursive} we develop a new procedure. It will be shown how the connected correlators \eqref{tp:wconn} can be constructed from the knowledge of the one-point functions $\mathcal{W}^{(k)}$ and develop a procedure by which the $1/N$ series can be construced. This will be the main new result of the paper.

\subsection{Explicit construction}
\label{tp:explicit}

Our starting point is the exact expression \eqref{ex:delAA},
\begin{equation}
\label{tp:delAA1}
	(\partial_1-\partial_2) \Tr(A_{z_1}A_{z_2}) = 
	 (z_1-z_2) \e{\frac{z_1^2+z_2^2}2}\sum\limits_{k=0}^{N-1}
	\Laguerre^{2k+1}_{N-1-k}(-z_1^2-z_2^2)
	\frac{(z_1z_2)^{2k}}{(k!)^2}~.
\end{equation}
Unfortunately, there does not appear to be an easy way to integrate \eqref{tp:delAA1} in this form, but we can proceed to expand it as we did with the one-point function in subsection~\ref{fund:explicit}. First, we use \cite[18.11.2]{NIST} to express the Laguerre polynomial in terms of a Whittaker function,
\begin{equation}
\label{tp:delAA2}
	(\partial_1-\partial_2) \Tr(A_{z_1}A_{z_2}) = 
	(z_1-z_2) \sum\limits_{k=0}^{\infty} \binom{N+k}{2k+1} \frac{(z_1z_2)^{2k}}{(-z_1^2-z_2^2)^{k+1}(k!)^2}  \WhitM_{N,k+\frac12}(-z_1^2-z_2^2)~,
\end{equation}
where we have formally extended the summation over $k$ to $\infty$, which is safe, because of the binomial coefficient. Then, expanding the Whittaker function into a series \cite[13.14.6]{NIST}, we get
\begin{align}
\label{tp:delAA3}
	(\partial_1-\partial_2) \Tr(A_{z_1}A_{z_2}) &= 
	(z_1-z_2) \sum\limits_{k=0}^{\infty} \sum\limits_{n=0}^\infty S(n,k;N) (z_1z_2)^{2k}
	\left(\frac{z_1^2+z_2^2}2\right)^n~,
\end{align}
where we have introduced the coefficients
\begin{equation}
\label{tp:S.1}
	S(n,k;N) =  
	\frac1{(k!)^2 n!} \binom{N+k}{2k+1} \genhypF{2}{1}{-n,k+1-N;2k+2;2}~.
\end{equation}

It is helpful to express \eqref{tp:delAA3} in terms of the variables 
\begin{equation}
\label{tp:zpm.def}
	z_\pm = \frac12(z_1\pm z_2)~,
\end{equation} 
and expand it in powers of $z_+^2-z_-^2$, which gives
\begin{equation}
\label{tp:delAA4}
	\partial_- \Tr(A_{z_1}A_{z_2}) = 2 z_- 
	\sum\limits_{k=0}^{\infty} \sum\limits_{n=0}^\infty S(n,k;N) \sum\limits_{j=0}^n \binom{n}{j} 
		(-1)^j (z_+^2-z_-^2)^{2k+j} (2z_+^2)^{n-j}~. 
\end{equation}
Equation \eqref{tp:delAA4} can be readily integrated. The result is
\begin{equation}
\label{tp:AA}
	\Tr(A_{z_1}A_{z_2}) = F(z_+) 
	- \sum\limits_{k=0}^{\infty} \sum\limits_{n=0}^\infty S(n,k;N) \sum\limits_{j=0}^n \binom{n}{j}(-1)^j
	\frac{(z_+^2-z_-^2)^{2k+j+1}(2z_+^2)^{n-j}}{2k+j+1}~.
\end{equation}
The integration constant $F(z_+)$ is determined uniquely by considering the special case $z_2=0$, in which $z_+=z_-$ and
\begin{equation}
\label{tp:F.spec}
	\Tr(A_{z_1}A_0)=\Tr A_{z_1} = \Tr A_{2z_+}~.
\end{equation}
Therefore, $F(z_+) = \Tr A_{z_1+z_2}$. Comparing this with \eqref{tp:wconn} reveals that the rest of \eqref{tp:AA} represents the connected two-point function,
\begin{equation}
\label{tp:W}
	\mathcal{W}^{(k_1,k_2)}_\text{conn} = \sum\limits_{k=0}^{\infty} \sum\limits_{n=0}^\infty S(n,k;N) \sum\limits_{j=0}^n \binom{n}{j}(-1)^j
		\frac{(z_+^2-z_-^2)^{2k+j+1}(2z_+^2)^{n-j}}{2k+j+1}~.
\end{equation}

Our next aim is to rewrite \eqref{tp:W} as a series in $1/N$. First, let us return to using $z_1$ and $z_2$, 
\begin{equation}
\label{tp:W2}
	\mathcal{W}^{(k_1,k_2)}_\text{conn} = 
	\sum\limits_{k=0}^{\infty} \sum\limits_{n=0}^\infty S(n,k;N)
	(z_1z_2)^{2k+n+1} \sum\limits_{j=0}^n \binom{n}{j} \frac{(-1)^j}{2k+j+1} \Delta^{n-j}~,
\end{equation}
where by $\Delta$ we denote the $N$- and $\lambda$-independent combination\footnote{The connected correlator vanishes when one of $k_1$ or $k_2$ vanishes.}
\begin{equation}
\label{tp:Delta}
	\Delta = \frac{(z_1+z_2)^2}{2z_1z_2} = \frac{(k_1+k_2)^2}{2k_1k_2}~.
\end{equation}
We show in appendix \ref{appS} that $S(n,k;N)$ has the form 
\begin{equation}
\label{tp:S.sum}
	S(n,k;N) = \sum\limits_{m=0}^{k+\left[\frac{n}2\right]} N^{2k+n+1-2m} \sigma(n,k,m)~.
\end{equation}
Therefore, substituting $z=kg$ and $g^2=\frac{\lambda}{4N}$ into \eqref{tp:W2} yields 
\begin{equation}
\notag 
	\mathcal{W}^{(k_1,k_2)}_\text{conn} = 
	\sum\limits_{k=0}^{\infty} \sum\limits_{n=0}^\infty \sum\limits_{m=0}^{k+\left[\frac{n}2\right]} N^{-2m} \sigma(n,k,m)
	\left(\frac{k_1k_2\lambda}{4}\right)^{2k+n+1} \sum\limits_{j=0}^n \binom{n}{j} \frac{(-1)^j}{2k+j+1} \Delta^{n-j}~.
\end{equation}
Then, pulling the sum over $m$ in front, we get 
\begin{equation}
\label{tp:W.N.series}
	\mathcal{W}^{(k_1,k_2)}_\text{conn} = \sum\limits_{m=0}^\infty N^{-2m} W_m^{(k_1,k_2)}~,
\end{equation}
where the coefficients are given by
\begin{align}
\notag
	W_m^{(k_1,k_2)}	&= 
	\sum\limits_{k=0}^{\infty} \sum\limits_{n=\max(0,2m-2k)}^\infty \sigma(n,k,m)
	\left(\frac{k_1k_2\lambda}{4}\right)^{2k+n+1} \sum\limits_{j=0}^n \binom{n}{j} \frac{(-1)^j}{2k+j+1} \Delta^{n-j}~,\\
\notag 
	&= \sum\limits_{n=0}^{\infty} \sum\limits_{k=0}^{m+\left[\frac{n}2\right]} \sigma(n+2m-2k,k,m)
		\left(\frac{k_1k_2\lambda}{4}\right)^{n+2m+1}\\
\notag
	&\quad \times \sum\limits_{j=0}^{n+2m-2k} \binom{n+2m-2k}{j} \frac{(-1)^j}{2k+j+1} \Delta^{n+2m-2k-j}~,\\
\label{tp:W3}
	&= 	\sum\limits_{n=0}^{\infty} \left(\frac{k_1k_2\lambda}{4}\right)^{n+2m+1} \sum\limits_{j=0}^{n+2m}   \frac{(-1)^j\Delta^{n+2m-j}}{(j+1)!(n+2m-j)!}  A(n,j,m)~.
\end{align}
Here, we have performed a sequence of sum rearrangements and introduced the coefficients
\begin{equation}
\label{tp:A.def}
	A(n,j,m) = \sum\limits_{k=0}^{\left[\frac{j}2\right]} \frac{j!(n+2m-2k)!}{(j-2k)!} \sigma(n+2m-2k,k,m)~.
\end{equation} 
Equation~\eqref{tp:W3} is our main result of this subsection. It provides an explicitly calculable expression for $W_m^{(k_1,k_2)}$ as a series in $\lambda$ and function of $k_1$ and $k_2$. The coefficients $\sigma(n,k,m)$ are given by double sums involving combinatorial functions, see \eqref{appS:sigma.1} or \eqref{appS:sigma.2}, which makes the whole result quite unwieldy except for the leading case $m=0$. Nevertheless, \eqref{tp:W3} can be used to check the series expansion of expressions of $W_m^{(k_1,k_2)}$ derived by other means.

\subsection{Leading term}
\label{tp:leading}

In the case of the leading term, $W_0^{(k_1,k_2)}$, the sum \eqref{tp:W3} simplifies significantly. Using \eqref{appS:sigma.m0}, the coefficient $A(n,j,0)$ becomes
\begin{align}
\notag 
	A(n,j,0) &= \sum\limits_{k} \frac{j!\,2^{n-2k}}{(j-2k)!(n+1)! (k!)^2}\\
\notag 
	&= \frac{2^{n-j}}{(n+1)!} \sum\limits_k \sum\limits_p \binom{j}{2k}\binom{2k}{k} \binom{j-2k}{p}~. 
\end{align}
Then, relabelling $p\to p-k$ and exchanging the order of summation gives 
\begin{align}
\notag 
	A(n,j,0) &= \frac{2^{n-j}}{(n+1)!} \sum\limits_p \sum\limits_k \binom{j}{p}\binom{j-p}{k} \binom{p}{k}\\
\notag 
	&= \frac{2^{n-j}}{(n+1)!} \sum\limits_p \binom{j}{p}^2 \\
\label{tp:A.m0}
	&=  \frac{2^{n-j}}{(n+1)!} \binom{2j}{j} = \frac{2^{n+j}}{(n+1)!j!} \left(\frac12\right)_j~. 
\end{align}
Substituting \eqref{tp:A.m0} into \eqref{tp:W3} yields
\begin{align}
\notag
	W_0^{(k_1,k_2)} &= \frac{k_1k_2\lambda}4 \sum\limits_{n=0}^\infty \frac{\left(\frac{\Delta}{2}k_1k_2 \lambda \right)^n}{n!(n+1)!}
	\sum\limits_{j=0}^n \frac{(-n)_j \left(\frac12\right)_j \left(\frac{2}{\Delta}\right)^j}{j! (j+1)!}\\
\label{tp:W0}
	&= \frac{k_1k_2\lambda}4 \sum\limits_{n=0}^\infty \frac{\left(\frac{\Delta}{2}k_1k_2 \lambda \right)^n}{n!(n+1)!}
	\hypF{-n,\frac12;2;\frac{2}{\Delta}}~.
\end{align}
Using hypergeometric function identities, this can be written in several equivalent forms. In particular,
\begin{align}
\label{tp:W0.1}
	W_0^{(k_1,k_2)} &= 
	\frac{k_1k_2\lambda}4 \sum\limits_{n=0}^\infty 
	\frac{\left(\frac{\Delta-2}{2}k_1k_2 \lambda \right)^n\left(\frac12\right)_n}{n!(2)_n(2)_n}
	\hypF{-n,\frac32;\frac12-n;\frac{\Delta}{\Delta-2}}\\
\label{tp:W0.2}
	&= 
	\frac{k_1k_2\lambda}4 \sum\limits_{n=0}^\infty 
	\frac{\left(\frac{\Delta}{2}k_1k_2 \lambda \right)^n\left(\frac32\right)_n}{n!(2)_n(2)_n}
	\hypF{-n,\frac12;-\frac12-n;\frac{\Delta-2}{\Delta}}~.
\end{align}
We remark that 
$$ \frac{\Delta}{\Delta-2} = \left( \frac{k_1+k_2}{k_1-k_2}\right)^2~. $$ 
The above expressions do not simplify further in terms of generalized hypergeometric series, except for the special cases $|k_1|=|k_2|$. Setting, without loss of generality, $|k_1|=|k_2|=1$, we have
\begin{align}
\label{tp:W0.s1}
	W_0^{(1,1)} &= \frac{\lambda}4 \genhypF{1}{2}{\frac32;2,2;\lambda}~,\\
\label{tp:W0.s2}
	W_0^{(1,-1)} &= -\frac{\lambda}4 \genhypF{1}{2}{\frac12;2,2;\lambda}~.
\end{align}

Let us compare these expressions with the results of Beccaria and Tseytlin \cite{Beccaria:2020ykg}. They have calculated the genus expansion of the correlators 
\begin{equation}
\label{tp:BT1}
	\vev{\Tr U \Tr U} = N^2 \genhypF{1}{2}{\frac32;2,3;\lambda} + \frac{\lambda}4 \genhypF{1}{2}{\frac32;2,3;\lambda} 
	+\frac{7\lambda^2}{192} \genhypF{1}{2}{\frac52;3,4;\lambda} +\cdots
\end{equation}
and
\begin{align}
\label{tp:BT2}
	\vev{\Tr U \Tr U^{-1}} &= N^2 \genhypF{1}{2}{\frac32;2,3;\lambda} - \frac{\lambda}4 \genhypF{1}{2}{\frac12;2,3;\lambda} 
	-\frac{\lambda^2}{192} \genhypF{1}{2}{\frac32;3,4;\lambda}\\
\notag 
	&\quad +\frac{\lambda^3}{2304} \genhypF{1}{2}{\frac52;4,5;\lambda} +\cdots~.
\end{align}
In order to find the connected contributions to \eqref{tp:BT1} and \eqref{tp:BT2}, we need to subtract $\vev{\Tr U}^2$. From \eqref{fund:W1.expand} and \eqref{fund:W.m} we have 
\begin{equation}
\label{tp:1pt}
	\vev{\Tr U} = N \genhypF{0}{1}{-;2;\frac{\lambda}4} + \frac{\lambda^2}{384 N} \genhypF{0}{1}{-;3;\frac{\lambda}4} +\cdots~.
\end{equation}
To square this, we can use the product formula \eqref{apphyp:product}, in which, for our parameters, the $\genhypF{2}{3}{}$'s simplify to $\genhypF{1}{2}{}$'s. Furthermore, one can use the contiguous function relations of the generalized hypergeometric functions, which we review in appendix~\ref{apphyp}. This gives 
\begin{equation}
\label{tp:1pt.square}
	\vev{\Tr U}^2 = N^2 \genhypF{1}{2}{\frac32;2,3;\lambda} + \frac{\lambda^2}{192}  \genhypF{1}{2}{\frac52;3,4;\lambda}  +\cdots~.
\end{equation}
Thus, after subtracting \eqref{tp:1pt.square} from \eqref{tp:BT1} and \eqref{tp:BT2} and using again the contiguous function relations of appendix~\ref{apphyp}, one finds \eqref{tp:W0.s1} and \eqref{tp:W0.s2}, respectively.
 
Another form of $W_0^{(k_1,k_2)}$ is \cite{Akemann:2001st, Okuyama:2018aij} 
\begin{equation}
\label{tp:Ake.explicit}
	W_0^{(k_1,k_2)} = \frac{\sqrt{\lambda}{k_1k_2}}{2(k_1+k_2)} \left[ \BesselI[0](k_1 \sqrt{\lambda}) \BesselI[1](k_2 \sqrt{\lambda}) + \BesselI[0](k_2 \sqrt{\lambda})\BesselI[1](k_1 \sqrt{\lambda}) \right]~.
\end{equation}
To prove the equivalence with our result, let us first expand the modified Bessel functions in \eqref{tp:Ake.explicit} into series. After some rearrangement of the two infinite sums one gets
\begin{equation}
\label{tp:Ake.series}
	W_0^{(k_1,k_2)} = \frac{\lambda k_1k_2}4 \sum\limits_{n=0}^\infty \frac{\left(\frac{\lambda}4 \right)^n}{n!(n+1)!} 
	\sum\limits_{j=0}^n \binom{n}{j} \binom{n+1}{j} \frac{k_1^{2j}k_2^{2(n-j)+1}+k_1^{2(n-j)+1}k_2^{2j}}{k_1+k_2}~.
\end{equation}

Thus, to show that \eqref{tp:W0} is equal to \eqref{tp:Ake.explicit}, we have to establish that
\begin{equation}
\label{tp:to.prove}
	\sum\limits_{j=0}^n \binom{n}{j} \binom{n+1}{j} \frac{k_1^{2j}k_2^{2(n-j)+1}+k_1^{2(n-j)+1}k_2^{2j}}{k_1+k_2}
	= (2\Delta k_1k_2)^n \hypF{-n,\frac12;2;\frac2\Delta}~.
\end{equation}
Consider first the left hand side of \eqref{tp:to.prove}. For simplicity, we shall omit the summation limits using the convention to sum over all possible non-zero summands. Using the identity
\begin{equation}
\label{tp:binom.ident}
	\binom{n+1}{j} = \binom{n}{j} +\binom{n}{j-1}
\end{equation}
and letting $j \to n-j$ in the term with even powers of $k_2$ in the numerator, we find
\begin{equation}
\label{tp:to.prove.lhs}
	\text{lhs.} = \sum\limits_j \left[ \binom{n}{j}^2 k_1^{2j} k_2^{2(n-j)} + \binom{n}{j} \binom{n}{j+1} k_1^{2j+1} k_2^{2(n-j)-1} \right]~.
\end{equation}
This can be written even shorter as 
\begin{equation}
\label{tp:to.prove.lhs.1}
	\text{lhs.} = \sum\limits_{m=0}^{2n} \binom{n}{\left[ \frac{m}2 \right]}  \binom{n}{\left[ \frac{m+1}2 \right]} k_1^{m} k_2^{2n-m}~.
\end{equation}

To manipulate the right hand side of \eqref{tp:to.prove}, we first use the hypergeometric function identity \cite[9.137.16]{Gradshteyn}, which leads to
\begin{equation}
\notag 
	\text{rhs.} = (2\Delta k_1 k_2)^n \left[ \hypF{-n-1,\frac12;1;\frac2{\Delta}} + \frac{n+2}{2\Delta} \hypF{-n,\frac32;3;\frac2\Delta} \right]~.
\end{equation}
By means of the quadratic transformation law \cite[9.134.2]{Gradshteyn} and recalling the definition of $\Delta$ \eqref{tp:Delta}, this is equal to 
\begin{align}
\notag 
	\text{rhs.} &= (2\Delta k_1 k_2)^n \left[ \left( 1+\frac{k_1}{k_2} \right)^{-2n-2} \hypF{-n-1,-n-1;1;\frac{k_1^2}{k_2^2}} \right. \\
\notag
	&\quad \left. + \frac{n+2}{2\Delta} \left( 1+\frac{k_1}{k_2} \right)^{-2n} \hypF{-n,-n-1;2;\frac{k_1^2}{k_2^2}} \right]~.
\end{align}
After writing out the hypergeometric series, this becomes
\begin{equation}
\notag
	\text{rhs.} = \frac1{(k_1+k_2)^2} \sum\limits_j \left[ \binom{n+1}{j}^2 k_1^{2j}k_2^{2(n+1-j)} 
	+ \binom{n}{j}\binom{n+2}{j+1} k_1^{2j+1}k_2^{2(n-j)+1} \right]~.
\end{equation}
Finally, with the help of the identity \eqref{tp:binom.ident} one can show that
\begin{equation}
\label{tp:to.prove.rhs}
	\text{rhs.} = \sum\limits_j \left[ \binom{n}{j}^2 k_1^{2j} k_2^{2(n-j)} + \binom{n}{j} \binom{n}{j+1} k_1^{2j+1} k_2^{2(n-j)-1} \right]~,
\end{equation}
which is just \eqref{tp:to.prove.lhs}. Thus, we have proven \eqref{tp:to.prove}.

\subsection{Recursive construction}
\label{tp:recursive}

The result \eqref{tp:W3} for the genus-$m$ contribution to the connected two-point function, although exact as a power series in $\lambda$, is extremely unwieldy. Beyond the leading order term, a general pattern is not apparent, and operations such as finding the large-$\lambda$ behaviour would require further work. Therefore, we shall abandon this explicit approach. Which alternatives do we have for making progress? A look at the one-point function can help. As explained in subsection~\ref{onept:DG}, the easiest way to find the genus expansion of the one-point function $\mathcal{W}^{(k)}$ is to use the differential equation \eqref{fund:W1.deq} to construct a recursive series of differential equations for $W_m$, \eqref{fund:Wm.deq}. Although these are second-order differential equations, the physically relevant solutions are unique once the leading order solution $W_0$ is taken as the start of the recursion. So, the question is whether a similar technique exists for the two-point functions. In this section we shall see that the answer to this question is indeed affirmative.

To start, let us return to the exact expression \eqref{ex:2pt.Okuyama}, 
\begin{equation}
\label{tp:2pt.Okuyama}
	(z_1+z_2)(\partial_{z_1}-\partial_{z_2}) \Tr(A_{z_1}A_{z_2}) 
	=  N \e{\frac{z_1^2+z_2^2}2} \left[ 
	\Laguerre_N^0(-z_1^2) \Laguerre_{N-1}^0(-z_2^2)
	-\Laguerre_N^0(-z_2^2) \Laguerre_{N-1}^0(-z_1^2) \right]~.
\end{equation}
$\Tr(A_{z_1}A_{z_2})$ differs from the connected correlator $\mathcal{W}^{(k_1,k_2)}_\text{conn}$ by a minus sign and an additional term that depends only on $k_1+k_2$, cf.\ \eqref{tp:wconn}. Therefore, \eqref{tp:2pt.Okuyama} implies that
\begin{equation}
\label{tp:2pt.Okuyama.W}
	(z_1+z_2)(\partial_{z_1}-\partial_{z_2}) \mathcal{W}^{(k_1,k_2)}_\text{conn}
	= - N \e{\frac{z_1^2+z_2^2}2} \left[ 
	\Laguerre_N^0(-z_1^2) \Laguerre_{N-1}^0(-z_2^2)
	-\Laguerre_N^0(-z_2^2) \Laguerre_{N-1}^0(-z_1^2) \right]~.
\end{equation} 
Let us also recall the exact one-point function \eqref{ex:Tr.A.result.2}
\begin{equation}
\label{tp:W.k.exact}
	\mathcal{W}^{(k)} = \e{\frac{z^2}2} \Laguerre^1_{N-1}(-z^2)~,
\end{equation}
which satisfies the differential equation 
\begin{equation}
\label{tp:1pt.deq}
	\left( \partial_z^2 +\frac3z \partial_z - z^2 -4N \right) \mathcal{W}^{(k)} = 0~.
\end{equation}
Equation \eqref{tp:1pt.deq} can be established either by direct calculation or by changing the independent variable in \eqref{fund:Wm.deq}. 
One can show by direct comparison with \eqref{tp:W.k.exact} and using some Laguerre polynomial identities that \eqref{tp:2pt.Okuyama.W} is nothing but
\begin{equation}
\label{tp:2pt.deq}
	(z_1+z_2)(\partial_{z_1}-\partial_{z_2}) \mathcal{W}^{(k_1,k_2)}_\text{conn}
	= - \frac1{2N} \left[ 2(z_1^2-z_2^2) + z_1^2 z_2 \partial_{z_2} - z_2^2 z_1 \partial_{z_1} \right] \Wprod~.
\end{equation}

To continue, let us rewrite \eqref{tp:1pt.deq} and \eqref{tp:2pt.deq} in terms of $k$, $k_1$ and $k_2$ as independent variables, recalling that $z=kg = k \sqrt{\frac{\lambda}{4N}}$. Therefore, 
\eqref{tp:1pt.deq} becomes
\begin{equation}
\label{tp:1pt.deq.k}
	\left( \partial_k^2 +\frac3k \partial_k - \lambda -\frac{k^2 \lambda^2}{16 N^2} \right) \mathcal{W}^{(k)} = 0~.
\end{equation}
Similarly, \eqref{tp:2pt.deq} takes the form
\begin{equation}
\label{tp:2pt.deq.k}
	(\partial_1-\partial_2) \mathcal{W}^{(k_1,k_2)}_\text{conn} 
	=  \frac1{N^2} \tDk{0} \Wprod~,
\end{equation}
where, here and henceforth, $\partial_n$ is a shorthand for $\partial_n\equiv \partial_{k_n}$, and $\tDk{0}$ denotes the operator 
\begin{equation}
\label{tp:tD.0}
	\tDk{0} = -\frac{\lambda}{8(k_1+k_2)} \left[ 2 ( k_1^2 -k_2^2) + k_1^2 k_2 \partial_2 - k_2^2 k_1 \partial_1 \right]~.
\end{equation}

Next, consider the operator
\begin{equation}
\label{tp:D.0}
	\Dk{0} = \frac{\lambda k_1 k_2}{8(k_1+k_2)} \left[ 2 ( k_1 +k_2) + k_1 k_2 (\partial_1 + \partial_2 ) \right]~.
\end{equation}
Applying $(\partial_1-\partial_2)$ from the left yields 
\begin{equation}
\label{tp:del.D.0}
	(\partial_1-\partial_2)	\Dk{0} = \frac{\lambda}{8(k_1+k_2)} \left[ -2 ( k_1^2 -k_2^2) + 4 k_2^2 k_1 \partial_1 - 4 k_1^2 k_2 \partial_2 
	+ k_1^2k_2^2   (\partial_1^2 - \partial_2^2 ) \right]~.
\end{equation}
When acting with this on $\Wprod$, one can use \eqref{tp:1pt.deq.k} to replace the second derivatives, which yields 
\begin{equation}
\label{tp:del.D.0.W}
	(\partial_1-\partial_2)	\Dk{0} \Wprod = \left( \tDk{0} -\frac1{N^2} \tDk{1} \right) \Wprod~,
\end{equation}
where we have introduced the new operator 
\begin{equation}
\label{tp:tD.1}
	\tDk{1} = - \frac{\lambda^3}{2^7} k_1^2 k_2^2 (k_1-k_2)~. 
\end{equation}
Clearly, \eqref{tp:2pt.deq.k} and \eqref{tp:del.D.0.W} imply that 
\begin{equation}
\label{tp:2pt.deq.k.1}
	(\partial_1-\partial_2) \mathcal{W}^{(k_1,k_2)}_\text{conn} 
	=  \left[ \frac1{N^2} (\partial_1-\partial_2) \Dk{0} +\frac1{N^4} \tDk{1} \right] \Wprod~.
\end{equation}

This suggests the following recursive procedure. Let $\Dk{n}$ and $\tDk{n}$ be operators independent of $N$ and containing at most first derivatives with respect to $k_1$ or $k_2$ (the only allowed second derivative is the mixed $\partial_1\partial_2$). They are defined in a recursive fashion by  
\begin{equation}
\label{tp:del.D.n.W}
	(\partial_1-\partial_2)	\Dk{n} \Wprod = \left( \tDk{n} -\frac1{N^2} \tDk{n+1} \right) \Wprod
\end{equation}
and by fixing the integration constant in $\Dk{n}$ such that $\mathcal{D}_n^{(k_1,0)}=\mathcal{D}_n^{(0,k_2)}=0$. In \eqref{tp:del.D.n.W}, second derivatives acting on $\Wprod$ are eliminated using \eqref{tp:1pt.deq.k}.
Then, we immediately have the solution\footnote{To rewrite \eqref{tp:2pt.sol} for the two-point correlator $\vev{\Tr U^{k_1} \Tr U^{k_2}}$, one can add a term with $n=-1$ and $\Dk{-1}=1$.} 
\begin{equation}
\label{tp:2pt.sol}  
	\mathcal{W}^{(k_1,k_2)}_\text{conn} = \sum\limits_{n=0}^\infty N^{-2-2n} \,\Dk{n} \Wprod~,
\end{equation} 
from which we can read off the $N^{-2m} $ coefficient 
\begin{equation}
\label{tp:2pt.sol.genus}  
	W^{(k_1,k_2)}_m = \sum\limits_{n=0}^m  \Dk{n} \sum\limits_{l=0}^{m-n} W_l^{(k_1)} W_{m-n-l}^{(k_2)}~.
\end{equation} 
It is reassuring to verify that the leading order term $W^{(k_1,k_2)}_0$ is just \eqref{tp:Ake.explicit}.

Let us flesh out this procedure. We start by writing
\begin{equation}
\label{tp:Dn.def}
	\Dk{n} = a_n \partial_1 \partial_2 + b_n^+(\partial_1+\partial_2) + b_n^-(\partial_1-\partial_2) +c_n~,
\end{equation} 
and 
\begin{equation}
\label{tp:tDn.def}
	\tDk{n} = \tilde{b}_n^+(\partial_1+\partial_2) + \tilde{b}_n^-(\partial_1-\partial_2) +\tilde{c}_n~,
\end{equation} 
where $a_n$, $b_n^\pm$, $c_n$, as well as $\tilde{b}_n^\pm$ and $\tilde{c}_n$ are functions of $k_1$, $k_2$ and $\lambda$. After inserting \eqref{tp:Dn.def} and \eqref{tp:tDn.def} into \eqref{tp:del.D.n.W} and eliminating the second derivatives by means of \eqref{tp:1pt.deq.k}, the terms of order $N^0$ give rise to the following system of equations,
\begin{subequations}
\label{tp:sys1}
\begin{align}
\label{tp:sys1.1}
	(\partial_1-\partial_2) a_n - \frac3{k_1}a_n +\frac3{k_2}a_n - 2b_n^- &=0~, \\
\label{tp:sys1.2}
	(\partial_1-\partial_2) b_n^+ +\frac3{2k_1k_2} [(k_1-k_2)b_n^+ - (k_1+k_2)b_n^-] &= \tilde{b}_n^+~,\\
\label{tp:sys1.3}
	(\partial_1-\partial_2) b_n^- +\frac3{2k_1k_2} [(k_1-k_2)b_n^- - (k_1+k_2)b_n^+] +c_n-\lambda a_n &= \tilde{b}_n^-~,\\
\label{tp:sys1.4}
	(\partial_1-\partial_2) c_n +2 \lambda b_n^- &= \tilde{c}_n~.
\end{align}
\end{subequations}
Moreover, the terms of order $N^{-2}$ determine the coefficients in $\tDk{n+1}$,
\begin{subequations}
\label{tp:next}
\begin{align}
\label{tp:next.1}
	\tilde{b}^+_{n+1} &= - \frac{\lambda^2}{32} a_n \left(k_1^2-k_2^2\right)~,\\
\label{tp:next.2} 	
	\tilde{b}^-_{n+1} &= \frac{\lambda^2}{32} a_n \left(k_1^2+k_2^2\right)~, \\
\label{tp:next.3}
	\tilde{c}_{n+1} &= -\frac{\lambda^2}{16} \left[\left(k_1^2-k_2^2\right)b_n^+ + \left(k_1^2+k_2^2\right)b_n^-\right]~.
\end{align}
\end{subequations}
The functions corresponding to $\tDk{0}$ \eqref{tp:tD.0} are 
\begin{equation}
\label{tp:tDk0.fs} 
	\tilde{b}^+_0 = -\frac{\lambda(k_1-k_2)}{16(k_1+k_2)}k_1k_2~,\qquad 
	\tilde{b}^-_0 = \frac{\lambda}{16}k_1k_2~,\qquad
	\tilde{c}_0 = -\frac{\lambda}4 (k_1-k_2)~,
\end{equation} 
and the first solution $\Dk{0}$ \eqref{tp:D.0} is given by
\begin{equation}
\label{tp:Dk0.fs} 
	a_0=0~,\qquad b^+_0 = \frac{\lambda k_1^2k_2^2}{8(k_1+k_2)}~,\qquad 
	b^-_0 = 0~,\qquad c_0 = \frac{\lambda}4 k_1k_2~.
\end{equation} 

In order to make progress for $n>0$, let us introduce the variables
\begin{equation}
\label{tp:yz.def}
	y = \lambda k_1 k_2 ~,\qquad z = \sqrt{\lambda} (k_1+k_2)~,
\end{equation}
and let 
\begin{equation}
\label{tp:hat.vars}
	a_n= \frac1{\lambda} \hat{a}_n~,\qquad 
	b_n^+= (k_1+k_2) \hat{b}_n^+~,\qquad
	b_n^-= (k_1-k_2) \hat{b}_n^-~,\qquad
	c_n = \hat{c}_n~,	
\end{equation}
where the new variables $\hat{a}_n$, $\hat{b}_n^\pm$ and $\hat{c}_n$ are functions of $y$ and $z$. The various factors of $\lambda$ serve the purpose of removing it from the system. Moreover, we note from \eqref{tp:next.1} and \eqref{tp:next.2} that
\begin{equation}
\label{tp:tb.rel}
	\left(k_1^2+k_2^2 \right) \tilde{b}^+_n + \left(k_1^2-k_2^2 \right) \tilde{b}^-_n =0~,
\end{equation} 
which can be used to form a homogeneous equation from \eqref{tp:sys1.2} and \eqref{tp:sys1.3} (for $n>0$ only). Then, the system \eqref{tp:sys1},  with the right hand sides determined by \eqref{tp:next}, can be transformed into the following recursive system for the hatted variables,
\begin{subequations}
\label{tp:sys2}
\begin{align}
\label{tp:sys2.1}
	\partial_y \hat{a}_n -\frac3{y} \hat{a}_n + 2 \hat{b}^-_n &=0~, \\
\label{tp:sys2.2}
	(4y-z^2) \partial_y \hat{b}^-_n +(2y-z^2) \partial_y \hat{b}^+_n -3\hat{b}^+_n -\hat{b}^-_n +\hat{c}_n -\hat{a}_n &=0~,\\
\label{tp:sys2.3}
	\partial_y \hat{b}^+_n -\frac3{2y} \left( \hat{b}^+_n -\hat{b}^-_n \right) &= \frac1{32}\hat{a}_{n-1}~, \\
\label{tp:sys2.4}
	\partial_y \hat{c}_n -2  \hat{b}_n^- &= \frac1{16}\left[ z^2\hat{b}_{n-1}^+ + (z^2-2y)\hat{b}_{n-1}^-\right]~.
\end{align}
\end{subequations}
The case $n=0$ is special. In that case, the right hand sides of \eqref{tp:sys2} are to be replaced by $0$, $\frac{y^2}{8z^2}$, $\frac{y}{16z^2}$ and $\frac14$, respectively. The start values are given by
\begin{equation}
\label{tp:rec.start}
	\hat{a}_0 = 0~,\qquad \hat{b}_0^+= \gamma_0\frac{y^2}{z^2}~,\qquad
	\hat{b}_0^-=0~, \qquad \hat{c}_0 = \gamma_0 2y \qquad \left( \gamma_0=\frac18 \right)~.  
\end{equation}

The system \eqref{tp:sys2} can be solved recursively for $n>0$. In each step, one must impose that $\hat{a}_n$, $\hat{b}_n^\pm$ and $\hat{c}_n$ vanish for $y=0$, which gives a unique solution and ensures that it corresponds to the connected 2-point function. This essentially implies that one can construct a particular solution of the inhomogeneous equation in the form of polynomials in $y$. More precisely, one can make a solution ansatz in which the functions $\hat{a}_n$, $\hat{b}_n^\pm$ and $\hat{c}_n$ are given by $y^3$ times polynomials of degree $n-1$, with coefficients that depend algebraically on $z$. Then, finding the coefficients amounts to solving a system of linear algebraic equations. This can be easily coded. The solutions until $n=4$, obtained with the help of \cite{sagemath}, are listed in table~\ref{tp:table}.

\begin{table}[ht]
\caption{Table of the coefficients $\hat{a}_n$, $\hat{b}_n^\pm$ and $\hat{c}_n$. The displayed expressions must be multiplied by $\gamma_n y^3$, where $\gamma_n=2^{-(5n+3)}/{\left(\frac32\right)_n}$.  \label{tp:table}}
\begin{tabularx}{\textwidth}{|X|l|l|l|l|}
\hline $n$ & $\hat{a}$ & $\hat{b}^+$ & $\hat{b}^-$ & $\hat{c}$ \\
\hline
\hline
1 & 1 & 0 & 0 & 1 \\
\hline 
2 & $4y-4(z^2+4)$ & $y+2$ & $-2$ & $-y$ \\
\hline 
3 & $ (z^2+24)y^2 -16(3z^2+20)y$  & $ 4 y^2 -(5 z^2 + 8) y$ & $-(z^2 + 24) y $ & $(z^2 - 4) y^2 $ \\
 & $+8(3z^4+40z^2+160)$ & $- 8(3 z^2 +20)$ & $+ 8(3 z^2 + 20)$ & $+ 4 (3 z^2 + 20)y $ \\
\hline
4 & $ 12 (z^2 + 16) y^{3}$ & $(z^2 + 24) y^3 $ & $-18(z^2+16)y^2$ & $3(z^2-8)y^3$\\
 & $ - 12 (z^4 + 60 z^2 + 480) y^{2} $ & $-18(3 z^2+16) y^2$ & $+12(z^4+60z^2$ & $-6 (z^4-24z^2$ \\
 & $+ 576 (z^4 + 20 z^2 + 112) y$ & $+36 (z^4+4z^2-32) y$ & $+480)y$ & $-288)y^2$ \\ 
 & $-192( z^6 +30 z^4 + 336 z^2 $ & $+288(z^4+20z^2+112)$ & $-288(z^4+20z^2$ & $-144(z^4+20z^2$\\
 & $ +1344)$ & $ $ & $+112)$ & $ +112)y$ \\
\hline
\end{tabularx}
\end{table}

\subsection{Special cases $|k_1|=|k_2|$}

For completeness, we shall provide the explicit expressions for a few subleading terms in the special cases $k_1=k_2$ and $k_1=-k_2$. In these cases, it is possible to simplify the general expressions that result from \eqref{tp:2pt.sol.genus} by applying the product formula \eqref{apphyp:product} and the contiguous function relations listed in appendix~\ref{apphyp}. As before, we can limit the discussion to $|k_1|=|k_2|=1$, because the general case can be recovered by rescaling $\lambda$. Without further details, the first sub-leading terms in the case $k_1=k_2$ are
\begin{equation}
\label{tp:Wpp.1}
	W_1^{(1,1)} = \frac{\lambda^3}{192} \genhypF{1}{2}{\frac{5}{2}; 3,4 ; \lambda}
	+ \frac{5\lambda^4}{12288}  \genhypF{1}{2}{\frac{7}{2} ; 4,5 ; \lambda}~,
\end{equation}
\begin{align}
\label{tp:Wpp.2}
	W_2^{(1,1)} &=  \frac{\lambda^{5}}{23040} \genhypF{1}{2}{\frac{7}{2} ; 5,6  ; \lambda } 
	+ \frac{217\lambda^{6}}{44236800} \genhypF{1}{2}{\frac{9}{2}; 6,7 ; \lambda} \\
\notag
	&\quad	+ \frac{23\lambda^{7}}{176947200} \genhypF{1}{2}{\frac{11}{2}; 7,8 ; \lambda}
	+\frac{77\lambda^{8}}{90596966400} \genhypF{1}{2}{\frac{13}{2}; 8,9 ; \lambda}~.
\end{align}
For $k_1=-k_2$, we have
\begin{equation}
\label{tp:Wpm.1}
	W_1^{(1,-1)} = -\frac{\lambda^{4}}{36864} \genhypF{1}{2}{\frac{5}{2}; 4,5 ; \lambda}~,
\end{equation}
\begin{equation}
\label{tp:Wpm.2}
	W_2^{(1,-1)} = - \frac{\lambda^{6}}{44236800} \genhypF{1}{2}{\frac{7}{2}; 4,7 ; \lambda}
	-\frac{\lambda^{8}}{3170893824000} \genhypF{1}{2}{\frac{11}{2}; 8,9 ; \lambda}~.
\end{equation}
As for the leading term, these results can be compared to the results of \cite{Beccaria:2020ykg} using the contiguous function relations of appendix~\ref{apphyp}.

%% file: conc.tex
\section{Conclusions}
\label{conc}

In this paper, we have discussed various aspects of $\frac12$-BPS Wilson loops in $\mathcal{N}=4$ SYM theory, focusing one exact results that can be obtained starting from the Gaussian matrix model representation. 

First, we have reviewed the formulation of general Wilson loop generating functions in the language of symmetric functions, which allows to use combinatorial tools to translate different basis representations into each other. We have generalized this formulation to two-loop generating functions, where by two-loop we mean loops running along two different contours. 

Second, this formalism has been applied to the generating functions of $\frac12$-BPS Wilson loops in $\mathcal{N}=4$ SYM theory. These Wilson loops have a circular contour, but can run along this contour in either direction, so that the generic two-loop case is needed for the most general treatment. We have considered the generating function of the correlators of multiply wound Wilson loops, $\mathcal{W}^{(k_1,\ldots,k_h)}$, which was introduced earlier by Okuyama, and provided a generalization of his result to all orders in combinatorial terms, cf.\ \eqref{N4:ccorr}. Specifically, the connected $h$-point function of multiply wound Wilson loops (with arbitrary orientation) is obtained in terms of the traces of products of certain matrices $A_k$.

Third, we have reviewed how the matrices $A_k$ can be reformulated in terms of harmonic oscillator quantum mechanics. The simplest results of the matrix model, in particular the one-point functions $\mathcal{W}^{(k)}=\Tr A_k$, arise in this formulation in a curiously elegant fashion. For the two-point functions, this approach results in an exact first-order differential equation, upon which one can build. Although we have not considered the three- and higher-point correlators in this paper, it is reasonable to believe that the harmonic oscillator formulation carries a lot of potential for further progress also in these cases.

Fourth, we have reviewed the Drukker and Gross expansion of the one-point function $\mathcal{W}^{(1)}$ as a series in $1/N^2$ and added two new approaches. We have shown that the entire series can be constructed also from a recursive system of differential equations, cf.\ \eqref{fund:Wm.deq}. Furthermore, the direct approach, in which the power series in $\lambda$ is reordered such as to give a series in $1/N^2$, results in several explicit expressions of the numerical (rational) coefficients of the Drukker and Gross series, which were originally defined only in terms of a recursion. 

Last, we have considered the $1/N^2$ expansion of the connected two-point functions, $\mathcal{W}^{(k_1,k_2)}_\text{conn}$, using two different approaches, both of which start from the exact result of the harmonic oscillator approach mentioned above. The direct approach of reordering the series in $\lambda$ into a series in $1/N^2$ results in an exact, although unwieldy, solution to this problem. However, we have also shown how the connected two-point functions, $\mathcal{W}^{(k_1,k_2)}_\text{conn}$, are related to the product of one-point functions, $\mathcal{W}^{(k_1)}\mathcal{W}^{(k_2)}$, and constructed a systematic procedure to calculate the series coefficients  $W_m^{(k_1,k_2)}$ in terms of the series coefficients $W_m^{(k_1)}$ and $W_m^{(k_2)}$, cf.\ \eqref{tp:2pt.sol.genus}. This construction is perhaps the main result of the paper. It is possible that this result is related to other methods that exploit the integrability of the Gaussian matrix model, such as the Toda integrability structure, and it would be very interesting to investigate this. Finally, a generalization of our construction to three- and higher-point functions is left for the future.

%% file: appS.tex
\section{Explicit forms of $S(n,k;N)$ and $\sigma(n,k,m)$}
\label{appS}

In this appendix, we will derive several forms of the coefficients $S(n,k;N)$ defined in \eqref{tp:S.1}. 
Using a hypergeometric function identity, \eqref{tp:S.1} can be also written as  
\begin{equation}
\label{tp:S.2}
	S(n,k;N) = \frac1{(k!)^2 n!} \binom{N+k}{2k+1} \frac{(N+k+1)_n}{(2k+2)_n}
	\genhypF{2}{1}{-n,k+1-N;-k-n-N;-1}~.
\end{equation}
Depending on which expression we start with, we shall find different, but non-trivially equivalent, results. This is similar to the expressions \eqref{fund:A.expl} and \eqref{fund:A.expl.2} for the coefficients $A(n,m)$. We shall start by considering \eqref{tp:S.2}, which will directly show that $S(n,k;N)$ has an expansion in $1/N^2$. Writing out the hypergeometric series in \eqref{tp:S.2} and simplifying gives
\begin{align}
\notag
	S(n,k;N) 
	&= \frac1{(k!)^2 n!} \binom{N+k}{2k+1} \frac{(N+k+1)_n}{(2k+2)_n}
	\genhypF{2}{1}{-n,k+1-N;-k-n-N;-1}	\\
\notag
	&= \frac{1}{(k!)^2n!}
	\sum\limits_{l=0}^n \binom{n}{l} \binom{N+k+n-l}{2k+n+1} \\
\label{appS:S.1}
	&= \frac{1}{(k!)^2n!(2k+n+1)!}
	\sum\limits_{l=0}^n \binom{n}{l}(N-k-l)_{2k+n+1}~, 
\end{align}
where we have let $l\to n-l$ in the last step. Next, we use \eqref{fund:Poch.expand} to get
\begin{align}
\label{appS:S.2}
	S(n,k;N) 
 	&=\sum\limits_{m=0}^{2k+n} N^{2k+n+1-m} \frac{1}{(k!)^2n!(2k+n+1)!} \sum\limits_{l=0}^n \binom{n}{l}  \\
\notag
 	&\quad \times \sum\limits_i (-1)^{m+k+l+i} s(k+l+1,i+1)s(n+k-l+1,2k+n-m-i+1)~,
\end{align}
where the sum over $i$ comprises all non-vanishing summands. Changing the summation indices by $l\to n-l$ and $i\to 2k+n-m-i$ returns the same summand with parity $(-1)^m$, which shows that terms with odd $m$ are absent. Therefore, we find 
\begin{equation}
\label{appS:S.series}
	S(n,k;N) = \sum\limits_{m=0}^{k+\left[\frac{n}2\right]} N^{2k+n+1-2m} \sigma(n,k,m)
\end{equation}
with
\begin{align}
\label{appS:sigma.1}
	\sigma(n,k,m) &= \frac{1}{(k!)^2n!(2k+n+1)!} \sum\limits_{l=0}^n \binom{n}{l} \\
\notag
&\quad \times
	\sum\limits_i (-1)^{k+l+i} s(k+l+1,i+1)s(n+k-l+1,2k+n-2m-i+1)~.
\end{align}
The special case $m=0$, which gives the leading order result, can be calculated in closed form. In this case, only the term with $i=l+k$ contributes in the sum on the second line of \eqref{appS:sigma.1}, so that one finds
\begin{equation}
\label{appS:sigma.m0}
	\sigma(n,k,0) = \frac{1}{(k!)^2n!(2k+n+1)!} \sum\limits_{l=0}^n \binom{n}{l} = \frac{2^{n}}{(k!)^2n!(2k+n+1)!}~.
\end{equation}

Another form of $\sigma(n,k,m)$ can be obtained by starting from \eqref{tp:S.1}, which gives
\begin{align}
\notag
	S(n,k;N) 
	&= \frac1{(k!)^2 n!} \binom{N+k}{2k+1} 
	\genhypF{2}{1}{-n,k+1-N;2k+2;2}	\\
\label{appS:S.3}
	&= \frac{1}{(k!)^2n!}
	\sum\limits_{l=0}^n \binom{n}{l} 2^l \frac{(N-k-l)_{2k+l+1}}{(2k+l+1)!}~. 
\end{align}
This time, however, we write $(N-k-l)_{2k+l+1}=(N+k-2k-l)_{2k+l+1}$ to find
\begin{align}
\label{appS:S.3.5}
	S(n,k;N) 
	&= \frac{1}{(k!)^2n!}
	\sum\limits_{l=0}^n \binom{n}{l} \frac{2^l}{(2k+l+1)!} \sum\limits_{i=0}^{2k+l+1} s(2k+l+1,i) (N+k)^i~.
\end{align}
Expanding the binomial $(N+k)^i$ and exchanging the order of summation yields
\begin{equation}
\label{appS:S.4}
 	S(n,k;N) 
 	= \frac{1}{(k!)^2n!}
 	\sum\limits_{l=0}^n \binom{n}{l} \frac{2^l}{(2k+l+1)!} \sum\limits_{j=0}^{2k+l+1} N^j 
 	\sum\limits_{i=0}^{2k+l+1-j} s(2k+l+1,i+j) \binom{i+j}{i} k^i~.
\end{equation}
Now, the sum over $l$ can be extended to $-2k-1$, because of the binomial coefficient $\binom{n}{l}$, after which we can exchange the sums over $j$ and $l$ to get
\begin{equation}
\label{appS:S.5}
 	S(n,k;N) 
 	= \frac{1}{(k!)^2n!} \sum\limits_{j=0}^{2k+n+1} N^j \sum\limits_{l=0}^{2k+n+1-j}
    \binom{n}{l+j-2k-1} \frac{2^{l+j-2k-1}}{(l+j)!}  
 	\sum\limits_{i=0}^{l} s(l+j,i+j) \binom{i+j}{i} k^i~.
\end{equation}
Note that the summand with $j=0$ can be omitted, because the first binomial would be non-zero only for $l\geq 2k+1$, but 
$$ \sum\limits_{i=0}^l s(l,i) k^i = (k-l+1)_l $$
vanishes for $l>k$. Moreover, letting $m=2k+n+1-j$, we know from above that the terms with odd $m$ vanish, although this is not evident here.  Omitting these terms, we find \eqref{appS:S.series} with
\begin{align}
\notag
	\sigma(n,k,m) &= \frac{1}{(k!)^2n!} \sum\limits_{l=0}^{2m} \binom{n}{l+n-2m} \frac{2^{l+n-2m}}{(l-2m+2k+n+1)!} \\
\notag
	&\quad \times
	\sum\limits_{i=0}^l s(l-2m+2k+n+1,i-2m+2k+n+1) \binom{2k+n+1-2m+i}{i} k^i\\
\label{appS:sigma.2}
	&= \frac{1}{(k!)^2} \sum\limits_{i=0}^{2m} k^{2m-i} \binom{2k+n+1-i}{2m-i} \sum\limits_{l=0}^{i} 
	\frac{2^{n-l}s(2k+n+1-l,2k+n+1-i)}{l!(n-l)!(2k+n+1-l)!}~,
\end{align}
where we have again exchanged the order of the sums. The special case \eqref{appS:sigma.m0} can be read off directly. I have checked using computer algebra \cite{sagemath} that \eqref{appS:sigma.1} and \eqref{appS:sigma.2} give the same values.

%% file: apphyp.tex
\section{Some properties of generalized hypergeometric functions}
\label{apphyp}

In this appendix, we will review some properties of generalized hypergeometric series, with special regard to relations between contiguous functions. The main sources of this material are \cite{Rainville:1960} and the earlier \cite{Rainville:1945}, as well as \cite{NIST}. 

For non-negative integers $p$ and $q$ and with coefficients 
\begin{equation}
\label{apphyp:ab}
	\mathbf{a} = ( a_1, \ldots, a_p)~, \qquad \mathbf{b} = ( b_1, \ldots, b_q)~,
\end{equation}
we define 
\begin{equation}
\label{apphyp:ab.Poch}
	(\mathbf{a})_n = \prod\limits_{i=1}^p (a_i)_n~, \qquad 	(\mathbf{b})_n = \prod\limits_{i=1}^q (b_i)_n~.
\end{equation}
Then, the generalized hypergeometric series is defined by
\begin{equation}
\label{apphyp:F.def}
	\genhypF{p}{q}{\mathbf{a};\mathbf{b};z} = \sum\limits_{n=0}^\infty \frac{(\mathbf{a})_n z^n}{(\mathbf{b})_n n!}~.
\end{equation} 

The series \eqref{apphyp:F.def} is convergent for all $z$ if $p\leq q$ and diverges for all $z$ in the case $p>q+1$. In the case $p=q+1$, it is convergent for $|z|<1$, divergent for $|z|>1$ and convergent for $|z|=1$ if $\re(\sum_i b_i - \sum_i a_i)>0$. If one of the coefficients $a_i$ is a negative integer or zero, the series terminates, in which case the above generic statements of divergence or convergence are irrelevant.
If an element of $\mathbf{a}$ coincides with an element of $\mathbf{b}$, then this pair of parameters can be omitted. For example,
\begin{equation}
\label{apphyp:reduction}
	\genhypF{p}{q}{a,a_2,\ldots,a_p;a,b_2,\ldots,b_q;z} = \genhypF{p-1}{q-1}{a_2,\ldots,a_p;b_2,\ldots,b_q;z}~.
\end{equation}

Some notable special cases are 
\begin{align}
\label{apphyp:spec.1}
	\genhypF{0}{0}{-;-;z} &= \e{z}~,\\
\label{apphyp:spec.2}
	\genhypF{1}{0}{a;-;z} &= (1-z)^{-a}~,\\
\label{apphyp:spec.3}
	\genhypF{0}{1}{-;b+1;\frac14 z^2} &= \Gamma(b+1) \left(\frac{z}2\right)^{-b} \BesselI[b](z)~.
\end{align}  

Amongst the few known product formulas, there is \cite[16.12.1]{NIST} 
\begin{equation}
\label{apphyp:product}
	\genhypF{0}{1}{-;a;z} \genhypF{0}{1}{-;b;z} = \genhypF{2}{3}{\frac{a+b}2,\frac{a+b-1}2;a,b,a+b+1;4z}~.
\end{equation}

The derivative of $\genhypF{p}{q}{}$ is 
\begin{equation}
\label{apphyp:deriv}
	\frac{\rmd}{\rmd z}\, \genhypF{p}{q}{\mathbf{a};\mathbf{b};z} 
	= \frac{\prod_i a_i}{\prod_i b_i}\, \genhypF{p}{q}{\mathbf{a}+1;\mathbf{b}+1;z}~,
\end{equation}
where by $\mathbf{a}+1$ we intend that every element of $\mathbf{a}$ is increased by unity.

Let 
\begin{equation}
\label{apphyp:theta.def}
	\theta = z \frac{\rmd}{\rmd z}~.
\end{equation}
Then, using $\theta z^k = k z^k$ one can easily derive that the generalized hypergeometric function satisfies the following differential equation of degree $q+1$,
\begin{equation}
\label{apphyp:diff.eq}
	\left[ \theta \prod\limits_{i=1}^q (\theta + b_i-1) - z \prod\limits_{i=1}^p (\theta + a_i) \right] \genhypF{p}{q}{\mathbf{a};\mathbf{b};z} = 0~.
\end{equation}   
Two generalized hypergeometric functions are said to be contiguous, if their parameters differ by integers. 
As in the case of the standard hypergeometric functions, there exist a number of linear relations between contiguous functions. The differential equation \eqref{apphyp:diff.eq}, together with \eqref{apphyp:deriv}, is an example, but there are others. Following \cite{Rainville:1945}, we shall introduce some shorter notation,
\begin{align}
\label{apphyp:F.short}
	\operatorname{F} &= \genhypF{p}{q}{\operatorname{a};\operatorname{b};z}~,\\
\label{apphyp:Fa.plus}
	\operatorname{F}(a_1\pm) &= \genhypF{p}{q}{a_1\pm 1,a_2,\ldots;\operatorname{b};z}~,\\
\label{apphyp:Fb.plus}
	\operatorname{F}(b_1\pm) &= \genhypF{p}{q}{\mathbf{a};b_1\pm 1,b_2,\ldots;z}~,
\end{align}
and so on. Then, one has
\begin{subequations}
\label{apphyp:F.deriv.rel}
\begin{align}
	(\theta + a_i)\operatorname{F} &= a_i \operatorname{F}(a_i+)~,\\
	(\theta + b_i-1)\operatorname{F} &= (b_i-1) \operatorname{F}(b_i-)~,
\end{align}
\end{subequations}
from which follow the contiguous function relations
\begin{subequations}
\label{apphyp:contig}
\begin{align}
	(a_i - a_j)	\operatorname{F} &= a_i \operatorname{F}(a_i+) - a_j \operatorname{F}(a_j+)~,\\
	(b_i - b_j)	\operatorname{F} &= (b_i-1) \operatorname{F}(b_i-) - (b_j-1) \operatorname{F}(b_j-)~,\\
	(a_i - b_j+1) \operatorname{F} &= a_i \operatorname{F}(a_i+) - (b_j-1) \operatorname{F}(b_j-)~.
\end{align}
\end{subequations}
Other relations can be found in \cite{Rainville:1960}, but will not be used here.